\documentclass[preprint,12pt,authoryear]{elsarticle}
\usepackage[latin1]{inputenc}
\usepackage{amsfonts,amssymb,amsmath}
\usepackage{graphicx}
\usepackage{color,multirow}
\usepackage{amsmath}
\usepackage{float}
\usepackage{amsthm}
\usepackage{amsfonts}
\usepackage{graphicx}
\usepackage{pdflscape}
\usepackage[UKenglish]{babel}
\usepackage{amssymb} 
\usepackage[labelfont={bf},textfont={it}]{caption}
\usepackage{pdfpages}
\usepackage[export]{adjustbox}

\makeatletter
\renewcommand*\env@matrix[1][*\c@MaxMatrixCols c]{%
 \hskip -\arraycolsep
 \let\@ifnextchar\new@ifnextchar
 \array{#1}}
\makeatother

\DeclareMathOperator*{\argmax}{arg\,max}
\numberwithin{equation}{section}

\graphicspath{{Figures/}}

\journal{Journal of Theoretical Biology}

\begin{document}

\begin{frontmatter} 

\title{A stochastic individual-based model to explore the role of spatial interactions and antigen recognition in the immune response against solid tumours
}
\author[]{FR Macfarlane\corref{cor1}}
\author[]{MAJ Chaplain}
\author[]{T Lorenzi}
\address{School of Mathematics and Statistics, University of St Andrews, St Andrews KY16 9SS, United Kingdom}
\cortext[cor1]{Corresponding author: \emph{Email address}: frm3@st-andrews.ac.uk }

\begin{abstract}
Spatial interactions between cancer and immune cells, as well as the recognition of tumour antigens by cells of the immune system, play a key role in the immune response against solid tumours. The existing mathematical models generally focus only on one of these key aspects. We present here a spatial stochastic individual-based model that explicitly captures antigen expression and recognition. In our model, each cancer cell is characterised by an antigen profile which can change over time due to either epimutations or mutations. The immune response against the cancer cells is initiated by the dendritic cells that recognise the tumour antigens and present them to the cytotoxic T cells. Consequently, T cells become activated against the tumour cells expressing such antigens. Moreover, the differences in movement between inactive and active immune cells are explicitly taken into account by the model. Computational simulations of our model clarify the conditions for the emergence of tumour clearance, dormancy or escape, and allow us to assess the impact of antigenic heterogeneity of cancer cells on the efficacy of immune action. Ultimately, our results highlight the complex interplay between spatial interactions and adaptive mechanisms that underpins the immune response against solid tumours, and suggest how this may be exploited to further develop cancer immunotherapies. 
\end{abstract}

\end{frontmatter}

\section{Introduction}
\label{sec:1}
The immune system is a collection of cells, structures and processes that work together to remove harmful foreign material from the body. Understanding the mechanisms which underpin the cellular immune response to solid tumours is crucial for the development of new immunotherapy techniques, which can strengthen the natural human immune response to support the successful treatment of cancer \citep{frankel2013,june2018,mellman2011,ribas2018}. 

Immune cells, specifically dendritic cells (DCs) and cytotoxic T lymphocytes (CTLs), detect and target solid tumours through recognising and processing the small peptides (\emph{i.e.} the antigens) that are expressed by tumour cells \citep{Messerschmidt2016}. The antigenic composition of solid tumours can be heterogeneous, whereby each cell within the tumour mass may have an antigen profile characterised by different expression levels of tumour antigens. On the surface of a tumour cell, antigens are integrated with major histocompatibility complex (MHC) molecules allowing the DCs to recognise the antigens and present them to the CTLs. The CTLs then become activated and subsequently express the corresponding antigen receptor, which enables them to interact with the tumour cells expressing that particular antigen at a sufficiently high level~\citep{Coulie2014}. Further interactions between CTLs and tumour cells can then trigger tumour cell death.

There are three distinct classes of tumour antigens: tumour associated antigens (TAAs), tumour specific antigens (TSAs) and cancer testis antigens (CTAs). These antigens are expressed, respectively, by: both normal and cancer cells, cancer cells only, and both cancer cells and human germ-line cells. Germ-line cells do not contain MHC molecules and, therefore, these cells cannot present antigens that can be recognised by T cells. This makes CTAs a viable target for immunotherapy as targeting these antigens implies a lower risk of autoimmune reactions~\citep{Coulie2014}. One specific group of CTAs is represented by the melanoma associated antigen (MAGE) genes~\citep{Boon2006,Connerotte2008,Muller2009,Urosevic2005}. Within this group, the MAGE-A family consists of eleven genes that are linked to poor prognosis in various types of cancer~\citep{Coulie2014,Hartmann2016,Zajac2017}. In general, the expression of MAGE-A genes promotes tumour progression by enhancing tumour cell division and reducing apoptosis~\citep{VanTongelen2017}. {{Some of the specific functions of MAGE-A genes have been discovered and include: inducing chemoresistance through a decrease in p53~\citep{marcar2010,monte2006,yang2007}, enhancing tumour proliferation~\citep{costa2007} and decreasing the efficacy of anti-tumour drugs~\citep{hartmann2013,hartmann2014}.}} The MAGE-A antigens are commonly expressed in melanomas \citep{Boon2006,Connerotte2008,Coulie2014,Urosevic2005}, oesophageal cancers \citep{Zajac2017}, lung, breast, prostate and colorectal carcinomas \citep{Coulie2014}, and head and neck cancers \citep{Hartmann2016,Muller2009}. Therefore, finding an effective way of targeting these antigens via immunotherapy may be beneficial to the treatment of multiple types of cancer~\citep{Zajac2017}. For instance, clinical trials have shown that targeting MAGE-A3 can be a successful treatment option \citep{Chinnasamy2011,Connerotte2008}. 

The antigen expression profiles of tumour cells can evolve over time due to epigenetic and genetic mechanisms. Amongst epigenetic mechanisms, spontaneous epimutations are `stochastic and heritable changes in gene expression that leave the sequence of bases in the DNA unaltered'~\citep{Oey2014}. Through spontaneous epimutations, the MAGE-A antigen expression levels of tumour cells can change over time, which may result in variability between the antigen profiles of histological samples from the same patient or between patients~\citep{Urosevic2005}. 

One potential cause of spontaneous epimutations is DNA methylation, whereby methylation of specific promoter regions of the gene represses transcription. Consequently, if the gene is methylated the corresponding protein will not be expressed. MAGE genes are methylated in normal cells, but they can be de-methylated, and thus expressed, in cancer cells~ \citep{Boon2006,Chalitchagorn2004,Chinnasamy2011,Muller2009,Zajac2017}. Demethylation of these genes can become more prominent during cancer progression, which suggests that demethylation of the MAGE genes may support tumour development \citep{Coulie2014}. 

Cell-cell interactions involved in the immune response to cancer depend upon the spatial position of both immune cells and tumour cells within the tumour micro-environment \citep{Chaplin2010,Messerschmidt2016}. In particular, as a result of tumour growth, cells within the tumour can become more exposed to immune action depending on their location in relation to other cells of the tumour micro-environment~\citep{Hanahan2011}. Moreover, stochastic antigenic variations can induce further spatial heterogeneity within the tumour \citep{Boon2006, Yarchoan2017}. Additionally, the movement of immune cells is dictated by the spatial distribution of tumour antigens within the tumour micro-environment~\citep{Boissonnas2007}. 

Mathematical models are a useful tool for simulating and investigating biological systems, and have been increasingly used to describe tumour antigen expression and tumour-immune interactions. Tumour antigen expression and the effects of epigenetic and genetic events have been modelled through differential equation models~\citep{Asatryan2016,Lorenzi2016,Johnston2007,Tomasetti2010} and cellular-automaton (CA) models~\citep{Bouchnita2017,Manem2014}. Traditionally, tumour antigen expression and recognition by the immune system have been implicitly modelled by tuning the rates of T cell recruitment, T cell proliferation or tumour cell removal~\citep{Arciero2004,Balea2014,Besse2018,DeBoer1985,dePillis2009,Kose2017,Mallet2006}. More recently, these processes have been explicitly captured by mathematical models formulated in terms of either ordinary differential equations \citep{Balachandran2017,dOnofrio2011,Luksza2017} or integro-differential equations \citep{Delitala2013a,Delitala2013,Kolev2013,Lorenzi2015}. However, these models rely on the assumption that cells are well-mixed and, as such, they do not reflect spatial aspects of tumour-immune competition. On the contrary, the spatial and temporal dynamics of tumour-immune interactions have been described through partial differential equation (PDE) models~\citep{Al-Tameemi2012,Matzavinos2004,Matzavinos2004b} and hybrid PDE-CA models~\citep{dePillis2006,Mallet2006}, but these do not take into account the antigen expression and recognition processes.

In light of these considerations, we present here a spatial individual-based model of tumour-immune competition that explicitly captures antigen expression and recognition. In our model, each cancer cell is characterised by an antigen profile which can change over time due to either epimutations or mutations. The immune response against the cancer cells is initiated by the DCs that recognise the tumour antigens and present them to the CTLs. Consequently, T cells become activated against the tumour cells which express such antigens. Moreover, exploiting the modelling strategies that we have previously developed~\citep{Macfarlane2018}, the differences in movement between inactive and active immune cells are explicitly taken into account. Computational simulations of this model clarify the conditions for the emergence of tumour clearance, dormancy or escape, and allow us to assess the impact of antigenic heterogeneity of cancer cells on the efficacy of immune action. Ultimately, our results highlight the complex interplay between spatial interactions and adaptive mechanisms that underpins the immune response against solid tumours, and suggest how this may be exploited to further develop cancer immunotherapies. 

The remainder of this work is organised as follows. In Section~\ref{sec:2}, we present the individual-based model and detail how each biological mechanism is mathematically described. In Section~\ref{sec:3}, we parametrise the model and present the results of computational simulations. In Section~\ref{sec:4}, we discuss the results obtained and highlight their biological implications along with potential further applications of this work.

\section{The mathematical model}
\label{sec:2}
Building upon our previous work~\citep{Macfarlane2018}, we consider three cell types: tumour cells, dendritic cells and cytotoxic T lymphocytes. We use an on-lattice individual-based approach to describe the interactions between these three cell types. Our model is posed on a 2D spatial grid of spacing $\Delta_{x}$ in the $x$ direction and $\Delta_{y}$ in the $y$ direction, with the constraint that only one cell of any type is allowed at each grid-site, at any time-step of duration $\Delta_{t}$.

The system is initially composed of cancer cells and inactive immune cells only. The immune cells are randomly distributed on the spatial grid, while the cancer cells are tightly packed in a circular configuration positioned at the centre of the grid, to reproduce the geometry of a solid tumour. We let the tumour grow through cell division. A cancer cell divides at rate $\lambda$ into two progeny cells of which one occupies the position of the parent cell while the other is positioned at an unoccupied neighbouring grid-site. This ensures that only cancer cells with free grid-sites in their neighbourhood can divide. Inactive immune cells update their position according to a L\'{e}vy-like walk~\citep{Harris2012}. This process allows them to move in a randomly chosen direction for a number of time-steps $s$ sampled from a L\'{e}vy distribution $L(s)$ with exponent $0 \leq \alpha <2$, \emph{i.e.} $L(s)\sim s^{-(\alpha+1)}$. DCs are activated at rate $D_{Act}$ upon contact with tumour cells, and CTLs become activated at rate $C_{Act}$ upon contact with active DCs. Upon activation, DCs and CTLs switch to Brownian motion, \emph{i.e.} at each time-step they can move to any of the neighbouring grid sites with the same probability. Moreover, we let active CTLs remove tumour cells, upon contact, at rate $\mu$. Once activated, both DCs and CTLs remain active throughout the simulations. For simplicity, we omit natural death of tumour cells and proliferation of both DCs and CTLs (\emph{i.e.} the total numbers of DCs and CTLs are constant over time). We refer the reader to our previous paper~\citep{Macfarlane2018} for a detailed description of these modelling strategies. 

\begin{figure}[h!]
\includegraphics[width=1.1\textwidth]{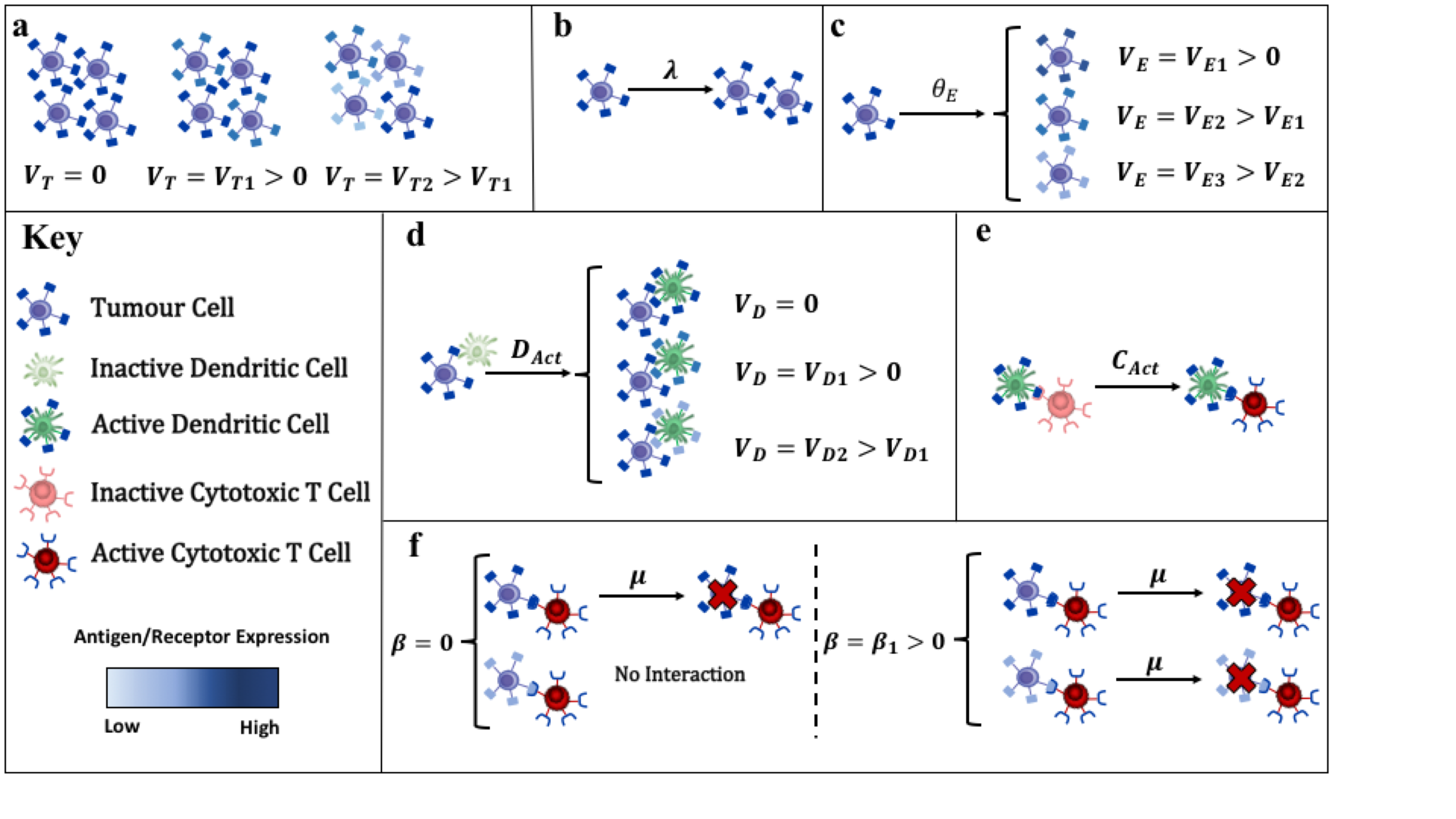}
\caption{\textbf{Schematic representation of the mechanisms and processes included in the individual-based model.}  We consider three cell types in the model: tumour cells, DCs and CTLs, along with their corresponding antigen and receptor profiles {{(\textbf{Key})}}. \textbf{a} Initially the tumour is composed of tumour cells characterised by different antigenic profiles. {{The standard deviation of the initial tumour antigen profiles from the reference experimental profile is given by the parameter $V_{T}$}}. \textbf{b} Tumour cells divide at rate $\lambda$. \textbf{c} Tumour cells may undergo epimutations with probability $\theta_{E}$. {{The standard deviation of the epimutation altered tumour antigen profiles from the previous profiles is given by the parameter $V_{E}$.}} \textbf{d} DCs become activated upon contact with tumour cells at rate $D_{Act}$. {{The standard deviation of the antigen profiles recognised by DCs from the tumour antigen profiles is given by the parameter $V_{D}$.}} \textbf{e} Upon contact, active DCs present the antigen profile they have recognised to inactive CTLs. This leads to the targeted activation of CTLs against specific tumour antigens at rate $C_{Act}$. \textbf{f} Activated CTLs remove tumour cells, upon contact, at rate $\mu$, on the condition that the tumour cells express a sufficient amount of the antigens corresponding to the CTL receptors. The binding affinity of the CTLs is measured by the parameter $\beta$.}
\label{fig:1} 
\end{figure}

In this work, we develop each of these modelling strategies further to include the antigen profiles of cancer cells and their possible variation, the immune recognition of tumour antigens by DCs, and the targeted activation of CTLs against specific tumour antigens. The modelling strategies used to take into account such additional layers of biological complexity are described in detail in the following subsections, and are also schematically illustrated in Figure~\ref{fig:1} and Figure~\ref{fig:2}.

\subsection{Mathematical modelling of antigen expression}
\label{sec:2_1}
We denote by $N_{T}(t)$ the number of tumour cells in the system at time $t = h \, \Delta_{t}$, with $h \in \mathbb{N}_0$, and we label each cell by an index $n=1,\ldots,N_T(t)$. We incorporate antigen expression into our model by letting each tumour cell express eleven different antigens, to represent the eleven MAGE-A antigens that, as mentioned in Section~\ref{sec:1}, have a key role in tumour development \citep{Coulie2014}. These antigens are reported in Table~\ref{tab:1} and we label them by an index $i=1,\ldots,11$. There can be high variability in each antigen's expression between patients with the same type of cancer \citep{Hartmann2016,Muller2009,Urosevic2005} and even within cancer cell samples from the same patient \citep{Hartmann2016,Muller2009,Urosevic2005}. Therefore, at each time instant $t$, we characterise the antigen profile of the $n^{th}$ tumour cell by means of a vector
$$
{\bf{A}}_{Tn}(t)=(A_{Tn}^{(1)}(t),\ldots,A_{Tn}^{(11)}(t)),
$$
with $A_{Tn}^{(i)}(t)$ representing the expression level of antigen $i$. {{Biologically, there is evidence that there can be correlation between antigen expression in some cancers, \emph{e.g.} epithelial ovarian cancer~\citep{daudi2014}, but not all cancers, \emph{e.g.} hepatocelllular carcinoma~\citep{roch2010}. To consider a more generalised situation, we assume that the expression levels of each antigen $i$ can evolve independently from the others.}} As schematically illustrated in Figure~\ref{fig:1}, for each tumour cell $n$ we define the initial expression of the $i^{th}$ antigen as
\begin{equation}
A_{Tn}^{(i)}(0)=(M_{i} + V_{T}\ R_i )_{+}, \quad i=1,\ldots,11. 
\label{eq:1}
\end{equation}
In equation~\eqref{eq:1}, the parameter $M_{i}$ denotes a mean expression level of antigen $i$ taken from published experimental data, the values of which are reported in Table~\ref{tab:1}.  {{The value of $R_i$ is sampled from a standard normal distribution centred at zero. In equation~\eqref{eq:1} we take the positive part of the right-hand side to ensure non-negativity of the antigen expression level. As $R_{i}$ is taken from a standard normal distribution, the parameter $V_{T}$ represents the standard deviation of the initial antigen profile from the experimental value $M_{i}$~\citep{stats}.}} Therefore, the parameter $V_{T}$ determines how close the value of $A_{Tn}^{(i)}(0)$ will be to the value of $M_{i}$.

\subsection{Modelling variations in antigen expression}
\label{sec:2_2}
\begin{figure}
\includegraphics[width=1\textwidth,frame=0.2mm]{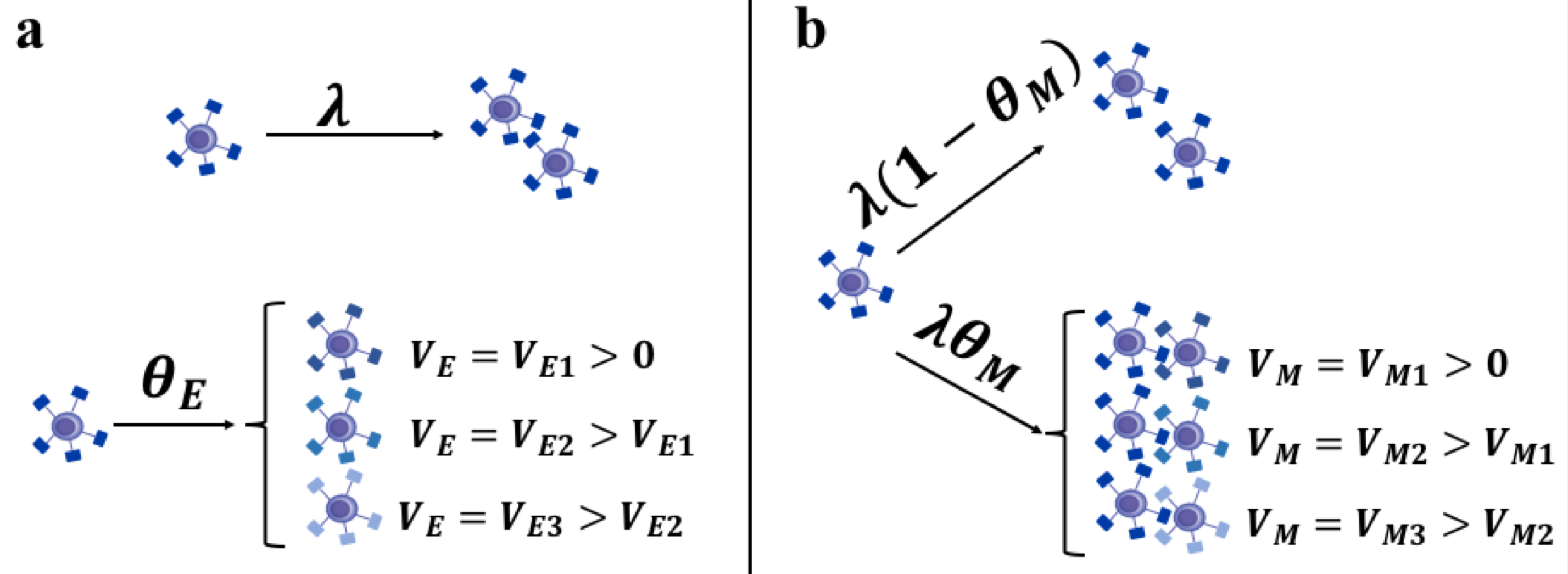}
\caption{\textbf{Schematic comparison of the modelling strategies used to describe changes in antigen expression induced by epimutations and mutations within tumour cells.} \textbf{a} Antigenic variations due to epimutations can occur, with probability $\theta_{E}$, at any time during the life of a tumour cell. {{The standard deviation of the new antigen profile from the previous profile is given by the parameter $V_{E}$.}} \textbf{b} Mutations can take place, with probability $\theta_{M}$, only during cell division, which occurs at rate $\lambda$. Due to mutations, one progeny cell may exhibit an antigen profile different from that of the parent cell. {{The standard deviation between the parent and progeny cell antigen profiles is given by the parameter $V_{M}$}}.} 
\label{fig:2} 
\end{figure}

At each time-step, we let the tumour cells divide at rate $\lambda$ [{{refer to}} the schemes in Figure~\ref{fig:1}\nobreak\hspace{0em}b] and change their antigen profile either through epimutations or through mutations. We assume that epimutations can occur at any time during the life of a cell [{{refer to}} the schemes in Figure~\ref{fig:1}\nobreak\hspace{0em}c and Figure~\ref{fig:2}\nobreak\hspace{0em}a], whereas mutations take place during cell division and may cause the antigen profile of one progeny cell to be different from that of the parent cell [{{refer to}} the scheme in Figure~\ref{fig:2}\nobreak\hspace{0em}b]. We allow epimutations and mutations to occur with probabilities $\theta_{E}$ and $\theta_M$, respectively. In the absence of changes in antigen expression (\emph{i.e.} if $\theta_E=0$ and $\theta_M=0$), the antigen profiles of the tumour cells will remain constant over time, that is, ${\bf{A}}_{Tn}(t)={\bf{A}}_{Tn}(0)$ for all $t>0$. If antigenic changes do occur through epimutations or mutations, the antigen profiles of tumour cells are updated using the methods described in the following paragraphs.

\paragraph{Epimutations} A variation in the level of expression of the $i^{th}$ antigen of the $n^{th}$ tumour cell at the time instant $t$ due to an epimutation is modelled according to the following equation 
\begin{equation}
A_{Tn}^{(i)}(t+\Delta_{t})=( A_{Tn}^{(i)}(t)+V_{E}\ R_i )_{+}, \quad i=1,\ldots,11.
\label{eq:2}
\end{equation}
{{In equation~\eqref{eq:2}, the value of $R_i$ is sampled from a normal distribution centred at zero and we take the positive part of the right-hand side to ensure non-negativity of the antigen expression level. Following on from the definition of the parameter $V_{T}$, since $R_{i}$ is taken from a standard normal distribution the parameter $V_{E}$ is the standard deviation of the updated antigen profile from the previous expression profile~\citep{stats}. Therefore, $V_{E}$ determines how close the value of $A_{Tn}^{(i)}(t+\Delta_{t})$ will be to the value of $A_{Tn}^{(i)}(t)$.}}

\paragraph{Mutations} Upon division at the time instant $t$, the $n^{th}$ tumour cell is replaced by two cells, one labelled by the index $n$ and the other one labelled by the index $N_T(t)+1$. If mutations do not occur, the progeny cells inherit the antigen profile of the parent cell, \emph{i.e.} $A_{Tn}^{(i)}(t+\Delta_{t})=A_{Tn}^{(i)}(t)$ and $A_{T\, N_T(t) +1}^{(i)}(t+\Delta_{t})=A_{Tn}^{(i)}(t)$. Conversely, if a mutation occurs, the antigen profile of the progeny cells will be given by the following equations 
\begin{equation}
A_{Tn}^{(i)}(t+\Delta_{t})=A_{Tn}^{(i)}(t), \quad i=1,\ldots,11
\label{eq:2M1}
\end{equation}
and
\begin{equation}
A_{T \, N_T(t) +1}^{(i)}(t+\Delta_{t})=( A_{Tn}^{(i)}(t)+V_{M}\ R_i )_{+}, \quad i=1,\ldots,11.
\label{eq:2M2}
\end{equation}
Equations~\eqref{eq:2M1} and \eqref{eq:2M2} rely on notation analogous to that of equation~\eqref{eq:2} {{and, therefore, $V_{M}$ is the standard deviation of the progeny antigen profiles from the parent antigen profiles~\citep{stats}. The value of $V_{M}$ determines how close the value of $A_{T \ N_T(t) +1}^{(i)}(t+\Delta_{t})$ will be to the value of $A_{Tn}^{(i)}(t)$.}}

\subsection{Activation of immune cells}
\label{sec:2_3}
\paragraph{Activation of DCs} We denote by $N_D$ the number of DCs, which we assume to be constant, and we label each DC by an index $k=1,\ldots,N_D$. Activation of DCs occurs, at rate $D_{Act}$, through contact with tumour cells. At any time instant $t$, the $k^{th}$ DC is characterised by a recognised antigen profile 
$$
{\bf{A}}_{Dk}(t)=(A_{Dk}^{(1)}(t),\ldots,A_{Dk}^{(11)}(t)).
$$
We let all DCs be initially inactive and thus assume
$$
A_{Dk}^{(i)}(0)=0, \quad i=1,\ldots,11
$$
for all $k=1,\ldots,N_D$. {{There is biological evidence supporting heterogeneity within the antigen presentation process where a less prevalent antigen may be recognised and presented by the DCs~\citep{boes2002,fehres2014,ljunggren1990}.  Additionally, it is known that the MAGE-A genes have similar homology~\citep{roch2010,Zajac2017} and, therefore, there is a potential that they could be mis-recognised as each other~\citep{Graff2002,linette2013,raman2016,schueler1998,tong2004}}}. As schematically described by Figure~\ref{fig:1}\nobreak\hspace{0em}d, we consider the case where there may be potential variation in the antigen recognition process. To capture this idea, upon activation through contact with the $n^{th}$ tumour cell at the time instant $t$, we assign the recognised antigen profile of the $k^{th}$ DC using the following equation
\begin{equation}
A_{Dk}^{(i)}(t+\Delta_{t})=(A_{Tn}^{(i)}(t)+V_{D}\ R_i)_{+}, \quad i=1,\ldots,11.
\label{eq:3}
\end{equation}
{{In equation~\eqref{eq:3}, the value of $R_{i}$ is sampled from a normal distribution centred at zero and we take the positive part of the right-hand side to ensure non-negativity of the antigen expression level. We can describe $V_{D}$ as the standard deviation of the antigen profile recognised by each DC from the actual tumour antigen profile~\citep{stats}. Therefore, $V_{D}$ determines how close the value of $A_{Dk}^{(i)}(t+\Delta_{t})$ will be to the value of $A_{Tn}^{(i)}(t)$. Following the method of our previous work~\citep{Macfarlane2018}, once activated, we let the DCs remain activated against their recognised tumour antigen profile.}}

\paragraph{Activation of CTLs} 
We denote by $N_C$ the number of CTLs, which we assume to be constant, and we label each CTL by an index $m=1,\ldots,N_C$. As schematically described by Figure~\ref{fig:1}\nobreak\hspace{0em}e, activation of CTLs occurs, at rate $C_{Act}$, through contact with activated DCs. At any time instant $t$, each CTL $m$ has a receptor profile 
$$
{\bf{A}}_{Cm}(t)=(A_{Cm}^{(1)}(t),\ldots,A_{Cm}^{(11)}(t)).
$$
We let all CTLs be initially inactive and thus assume
$$
A_{Cm}^{(i)}(0)=0, \quad i=1,\ldots,11
$$
for all $m=1,\ldots,N_C$. While DCs can recognise multiple types of antigens, CTLs can produce copies of one antigen receptor only~\citep{Brenner2008,Coico2015a}. This means that each CTL can only be activated against one of the eleven MAGE-A antigens. To capture this fact, upon activation through contact with the $k^{th}$ DC at the time instant $t$, we let the $m^{th}$ CTL become activated against the highest expressed antigen within the tumour antigen profile recognised by the $k^{th}$ DC, \emph{i.e.} we assign the receptor profile of the $m^{th}$ CTL using the following equation
\begin{equation}
A_{Cm}^{(i)}(t)=
\begin{cases}
1 \mbox{ for } i= \hat{i}, \\
0 \mbox{ for } i \ne \hat{i},
\end{cases}
\quad \text{with } \; \hat{i} = \argmax_{j} A_{Dk}^{(j)}(t),
\label{eq:4}
\quad
\end{equation}
where the index $\hat{i}$ specifies the target antigen of the activated CTL. Note that, if $|\argmax_{j} A_{Dk}^{(j)}(t)| > 1$, then we arbitrarily choose $\hat{i} = \min \argmax_{j} A_{Dk}^{(j)}(t)$. Using the same assumptions as in our previous work, once activated, a CTL remains activated against the same tumour antigen.

\subsection{Removal of tumour cells by activated CTLs}
Upon contact, each activated CTL can induce death of the tumour cells which express a sufficiently high level of the CTL's target antigen~\citep{Coulie2014,Stone2009}, which we assume to be given by the mean antigen expression levels reported in Table~\ref{tab:1}. In particular, as schematically described by Figure~\ref{fig:1}\nobreak\hspace{0em}f, when the $m^{th}$ CTL interacts with the $n^{th}$ tumour cell at time $t$, we compare the receptor profile ${\bf{A}}_{Cm}(t)$ with the antigen profile ${\bf{A}}_{Tn}(t)$ and let the tumour cell be removed from the system at rate $\mu$ provided that
\begin{equation}
A_{Tn}^{(i)}(t)\ge (M_{i}-\beta) \; \mbox{ for } \; i \; \text{ such that } \; A_{Cm}^{(i)}(t)=1.
\label{eq:5}
\end{equation}
In equation \eqref{eq:5}, the parameter $\beta$ describes the binding affinity of the CTLs, which determines the range of tumour cells that each CTL can interact with. If $\beta$ is larger, then the CTL can recognise tumour cells with a lower level of expression of the antigen that they target. Independently of the outcome of the interaction, the CTL can subsequently interact with further tumour cells following the same process.

\section{Computational simulations}
\label{sec:3}
\subsection{Model parametrisation and simulation set up}
\label{sec:3_1}
We use a 2D spatial domain with $100$ grid sites, of length $\Delta_{x}=\Delta_{y}=10\ \mu$m, both in the $x$ and in the $y$ direction, which correspond to a domain of size $1\ \text{mm}^{2}$. Simulations were developed and run in {\sc Matlab}, for an appropriate number of time-steps, with one time-step chosen to be $\Delta_{t}=1$ min, to allow for the resulting dynamics of the system to be investigated. All quantities in the results we report on in this section were obtained through averaging the results of $5$ simulations. We refer the interested reader to \cite{Macfarlane2018} for a detailed description of the parameterisation of the original model, and we describe here the way in which the additional components of the model were calibrated using the parameter values reported in Table~\ref{tab:2}.

\begin{table}
\caption{Model parameters and related values used in computational simulations. Note, standard deviation has been abbreviated to StD.}
\label{tab:2} 
\resizebox{\textwidth}{!}{
\begin{tabular}{llll}
\hline\noalign{\smallskip}
Symbol & Description & Value(s) & Reference \\
\noalign{\smallskip}\hline\noalign{\smallskip}
$\Delta_{t}$ & time-step & $1$ min & \citep{Boissonnas2007} \\ 
$\Delta_{x, y}$ & grid spacing in the $x$ or $y$ direction & $10\ \mu $m & \citep{Macfarlane2018} \\
$N_{T}(0)$ & initial number of tumour cells & $400$ cells & \citep{Christophe2015}\\
$N_{C}$ & total number of CTLs & $400$ cells & \citep{Christophe2015}\\
$N_{D}$ & total number of DCs & $400$ cells & \citep{Macfarlane2018}\\
$n$ & Index identifier of each tumour cell &$n=1,\dots,N_{T}$&-\\
$k$ & Index identifier of each DC  &$k=1,\dots,N_{D}$&-\\
$m$ & Index identifier of each CTL &$m=1,\dots,N_{C}$&-\\
${\bf{A}}_{Tn}(t)$ & Antigen profile of tumour cell $n$ at time $t$ &values $\ge 0$& -\\
${\bf{A}}_{Dk}(t)$ & Recognised antigen profile of DC $k$ at time $t$ &values $\ge 0$& -\\
${\bf{A}}_{Cm}(t)$ & Antigen receptor profile of CTL $m$ at time $t$ &values of 0 or 1& -\\
$\lambda$ & tumour cell division rate & $0.001\ \text{min}^{-1}$ & \citep{Christophe2015}\\
$\theta_{E}^{*}$ & Average probability of epimutations & $0.23$ & \citep{DeSmet1996}\\
$\mu$ & removal rate of tumour cells by CTLs & $0.03\ \text{cells}\ \text{min}^{-1}$ & \citep{Christophe2015}\\
$D_{Act}$ & DC activation rate & $0.07\ \text{cells}\ \text{min}^{-1}$ & \citep{Bianca2012}\\
$C_{Act}$ & CTL activation rate & $\approx 0.12\ \text{cells}\ \text{min}^{-1}$ & \citep{Engelhardt2012}\\
$\alpha $ & L\'{e}vy walk exponent & 1.15 & \citep{Harris2012}\\
$\beta $ & T cell binding affinity &  $[0,0.2]$& \citep{schmid2010}\\
$V_{T} $ & StD of the initial tumour antigen profiles from the reference experimental profile {\bf $M$ } &   $[0,1]$ & -\\
$V_{D}$ & StD of the antigen profiles recognised by DCs from the tumour antigen profiles &  $[0,1]$ & -\\
$V_{E/M} $ & StD of the tumour antigen profiles from the previous profiles after epimutations/mutations & same as $V_{T}$ & -\\
\noalign{\smallskip}\hline
\end{tabular}}
\end{table}

\paragraph{Initial tumour antigen expression levels}
\cite{Hartmann2016} investigated the levels of expression of the eleven MAGE-A antigens in oral squamous cell cancers of $38$ patients. The mean antigen expression levels were taken to be values between $0$-$12$ arbitrary units as an immune-reactivity score, which are normalised and reported in Table~\ref{tab:1}. 
\begin{table}
\caption{Average levels of expression of the MAGE-A antigens in oral squamous cell cancer cell lines. The experimental data were taken from~\citep{Hartmann2016} and then normalised.} 
\label{tab:1} 
\centering
\begin{tabular}{cc} 
 \hline
 Antigen $(i)$ & Mean Expression, $M_{i}$ \\ 
\hline
 MAGE-A1 (1) & 0.10 \\
 MAGE-A2 (2) & 0.25 \\ 
 MAGE-A3 (3) & 0.41 \\ 
 MAGE-A4 (4) & 0.24 \\ 
 MAGE-A5 (5) & 0.36 \\ 
 MAGE-A6 (6) & 0.35 \\ 
 MAGE-A8 (7) & 0.17\\ 
 MAGE-A9 (8) & 0.16 \\ 
 MAGE-A10 (9) & 0.38 \\ 
 MAGE-A11 (10) & 0.06 \\ 
 MAGE-A12 (11) & 0.32\\ 
 \hline
\end{tabular}
\end{table}
In our model, the initial expression level of each antigen for each tumour cell is defined by using equation~\eqref{eq:1} along with the values of $M_{i}$ from Table~\ref{tab:1}. {{To determine the value of the product $V_{T}\ R_{i}$ in equation~\eqref{eq:1} we consider the properties of $R_{i}$, which is a random value taken from a standard normal distribution. A standard normal distribution $\mathcal{N}(0,1)$, with mean 0 and standard deviation 1, has a 95$\%$ confidence interval of $\pm$1.96~\citep{stats}. Therefore, for 95$\%$ of values we expect -1.96 $\le  R_{i} \le$1.96 with the majority of the values being close to the mean value. We then use the parameter $V_{T}$ to control the minimum and maximum value of the product $V_{T}\ R_{i}$, \emph{i.e.} when $V_{T}$=1 then $V_{T}\ R_{i}\in [-1.96,1.96]$, for most values. However, if $V_{T}$ is lower, \emph{e.g.} $V_{T}$=0.1, then the values of the product $V_{T}\ R_{i}$ will also be lower, \emph{e.g.} $V_{T}\ R_{i} \in [-0.196,0.196]$. To consider a wide range of biological situations corresponding to different initial levels of heterogeneity in tumour antigen profiles, we use a range of values, between 0 and 1, for the parameter $V_{T}$. As the experimental data in Table~\ref{tab:1} are dimensionless, we consider also $V_{T}$ to be dimensionless.}}
\paragraph{Probabilities of epimutations and mutations}
\cite{DeSmet1996} found that cancer cell lines expressing the MAGE-A1 antigen were $23\%$ more likely to undergo demethylation events than tumour cell lines that did not express this antigen. Such a value is supported by other studies that consider the likelihood of DNA demethylation in various cancers \citep{Chalitchagorn2004,Ehrlich2002}. We make the assumption that this holds true for the other ten MAGE-A antigens, and let the average probability of epimutations in our model be
$$
\theta_{E}^{*}=0.23.
$$
In the model we consider the effect of increase or decreasing the probability of epimutations or mutations by setting $\theta_{E}$ and $\theta_{M}$, respectively, as multiples of $\theta_{E}^{*}$. {{The parameters $V_{E}$ and $V_{M}$ control how much the antigen expression of a tumour cell can change through epimutations or how much the antigen expression of a progeny tumour cell can change through mutations [{{refer to}} equations~\eqref{eq:2} and~\eqref{eq:2M1}]. The values of $V_{M}$ and $V_{E}$ are chosen to match the values of $V_{T}$. Since the experimental data in Table~\ref{tab:1} are dimensionless, we also consider $V_{E}$ and $V_{M}$ to be dimensionless.}}.

\paragraph{Antigen recognition process:}
{{To consider a wide range of biological situations corresponding to different scenarios in terms of the number of CTLs activated against each antigen, we use a range of values, between 0 and 1, for the parameter $V_{D}$ [{{refer to}} equation~\eqref{eq:3}]. As the experimental data in Table~\ref{tab:1} are dimensionless, we also consider $V_{D}$ to be dimensionless.}}

\paragraph{T cell binding affinity:}
{{The binding affinity of a T cell is related to the association rate, that is the inverse of the dissociation rate $K_{D}$. In general this value is between $0.005\ \mu M^{-1}$ and $1\ \mu M^{-1}$ for all natural T cells~\citep{davis1998,slansky2010}. Furthermore, the MAGE-A T cell receptors association rates have been found to be even larger than this range, \emph{e.g.} for MAGE-A3 the values are between $0.018\ \mu M^{-1}$ and $5.917\ \mu M^{-1}$~\citep{Tan2015}. However, \cite{schmid2010} have shown that an association rate of $0.2\ \mu M^{-1}$ or higher did not improve the binding affinity and, therefore, higher binding affinities may have a limited effect. We take $M_{i}$ to be the minimal binding and allow the likelihood of binding to increase depending on $\beta$. In line with experimental evidence, we investigate a range of dimensionless values between $0$ and $0.2$ for the parameter $\beta$ that models the T cell receptor binding affinity [{{refer to}} equation~\eqref{eq:5}].}}

\subsection{Main Results}
\paragraph{Variability in the initial tumour antigen profiles determines the effectiveness of the immune response}
\label{sec:3_2}
\begin{figure}
\includegraphics[width=0.5\textwidth]{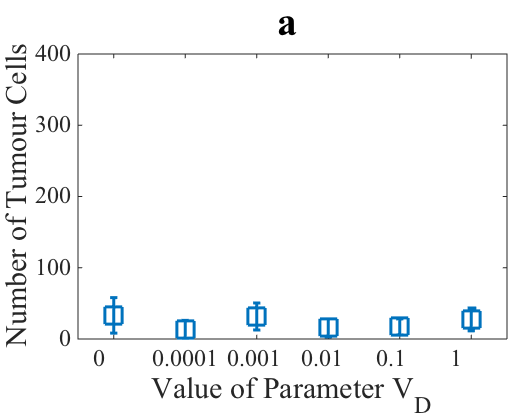} \includegraphics[width=0.5\textwidth]{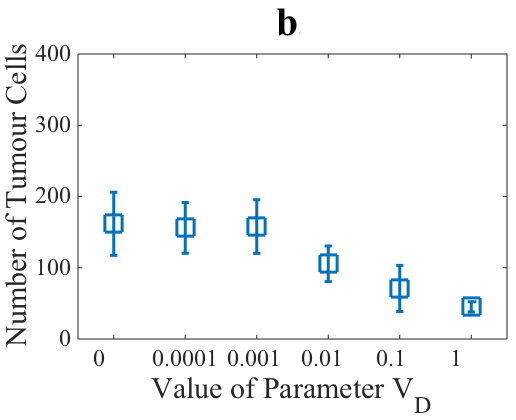}\\ \begin{center}\includegraphics[width=0.5\textwidth]{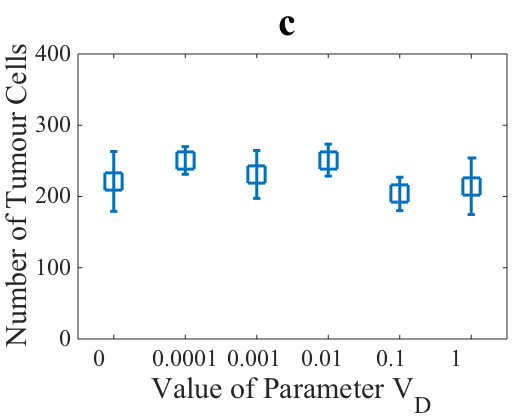}\end{center}
\caption{\textbf{Variability in the initial tumour antigen profiles determines the effectiveness of the immune response.} Plots displaying the number of tumour cells remaining after 1000 time-steps for increasing values of the parameter $V_{T}$: \textbf{a} $V_{T}=0.001$, \textbf{b} $V_{T}=0.01$ and \textbf{c} $V_{T}=0.1$. For each value of $V_{T}$ a range of values of the parameter $V_{D}$ are tested. The tumour cell numbers presented have been obtained as the average over $5$ simulations and the error bars display the related standard deviation. Here, $\beta=0.01$, $\theta_{E}=\theta_{M}=0$, and all the other parameter values are reported in Table~\ref{tab:2}.} 
\label{fig:3} 
\end{figure}

To investigate how the immune response is affected by variability in the initial tumour antigen profiles, we test for three increasing values of the parameter $V_{T}$ (\emph{i.e.} $V_{T}=0.001$, $V_{T}=0.01$ and $V_{T}=0.1$). We choose $\beta=0.01$ and we let the antigen profiles of the tumour cells remain constant over time (\emph{i.e.} we choose $\theta_{E}=\theta_{M}=0$). For each value of $V_{T}$ considered, we also explore the effect of increasing the value of the parameter $V_{D}$. In all cases, we carry out numerical simulations for $1000$ time-steps. As shown by Figure~\ref{fig:3}\nobreak\hspace{0em}a, for a low value of $V_{T}$, very few tumour cells remain in the system after $1000$ time-steps for all considered values of $V_{D}$. Conversely, the results presented in Figure~\ref{fig:3}\nobreak\hspace{0em}c show that if we set a relatively large value for $V_{T}$, there is a significant number of remaining tumour cells after $1000$ time-steps for all values of $V_{D}$. Moreover, as shown by Figure~\ref{fig:3}\nobreak\hspace{0em}b, for an intermediate value of $V_{T}$, there appears to be a correlation between the number of tumour cells remaining after $1000$ time-steps and the parameter $V_{D}$. In particular, larger values of the parameter $V_{D}$ correspond to smaller numbers of the remaining tumour cells after $1000$ time-steps. These results suggest that for tumours characterised by intermediate levels of initial antigenic heterogeneity between tumour cells, higher {{deviation}} between the antigen profile recognised by DCs and the actual antigen profile of tumour cells may result in a more effective immune response. This is further illustrated by the computational results presented in the next paragraph.

\paragraph{Increasing variations between the antigen profile recognised by DCs and the actual antigen profile of tumour cells can result in immune escape, chronic dormancy or immune clearance of the tumour}
\label{sec:3_3}
\begin{figure}
\centering
\includegraphics[height=0.32\textwidth]{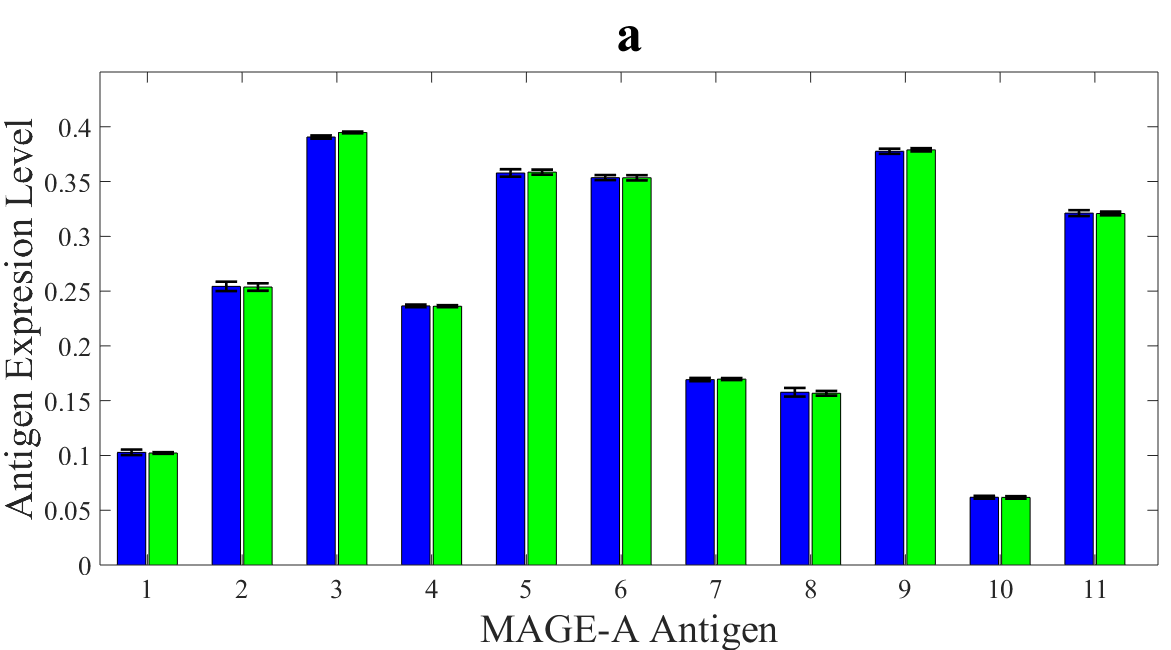}\ \includegraphics[height=0.32\textwidth]{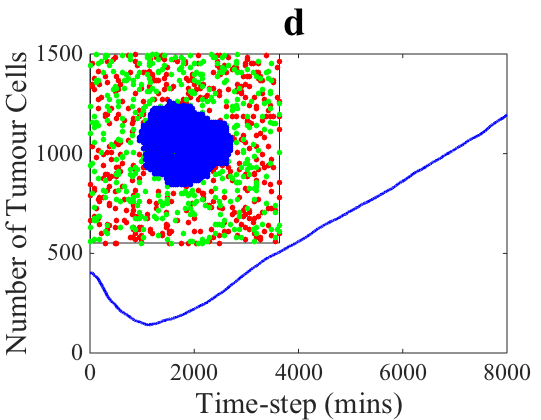}\\
\includegraphics[height=0.32\textwidth]{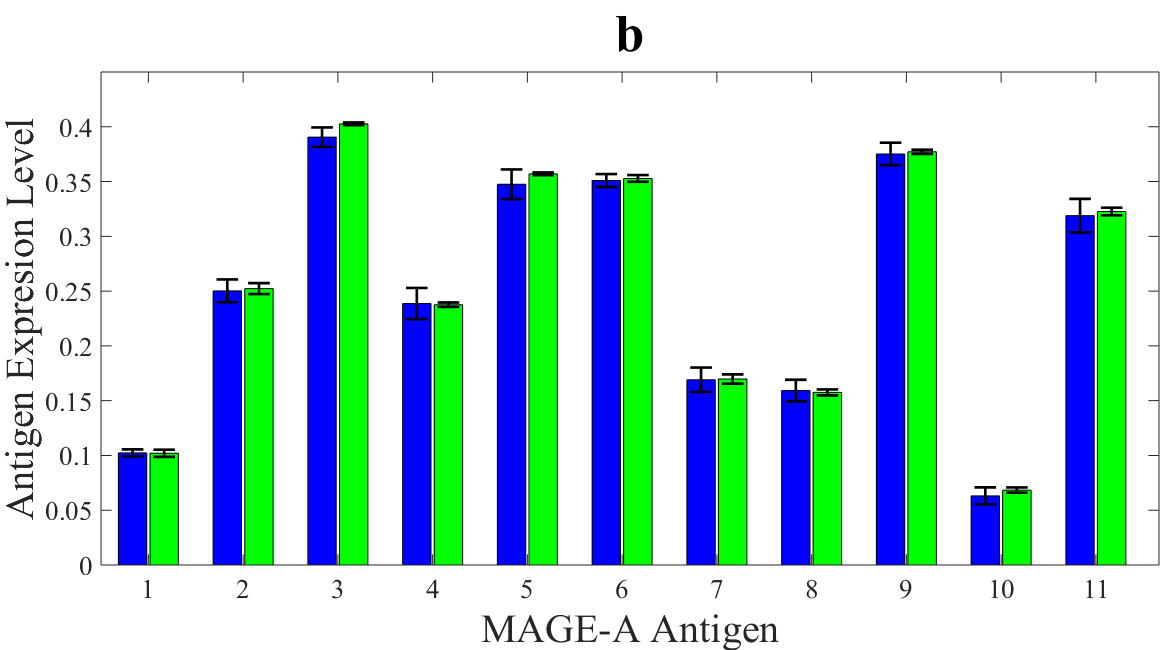}\ \includegraphics[height=0.32\textwidth]{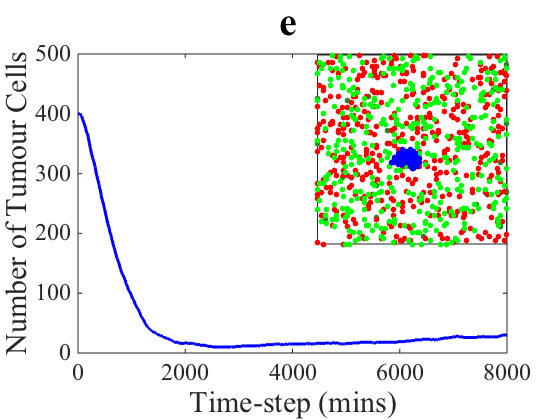}\\
\includegraphics[height=0.32\textwidth]{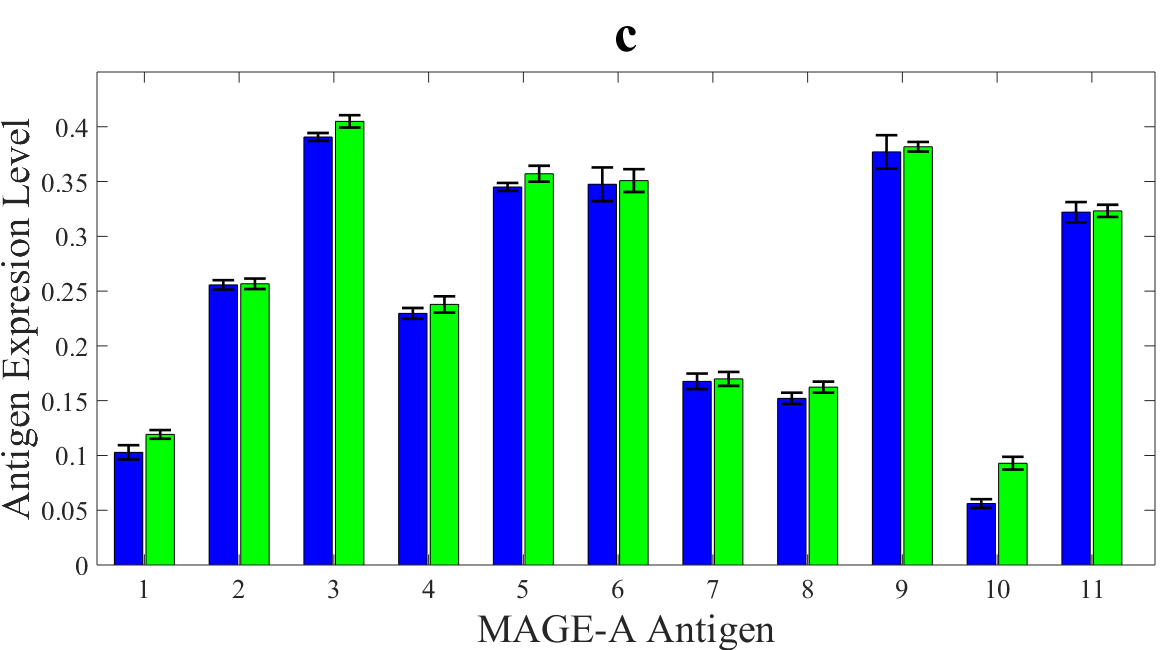}\ \includegraphics[height=0.32\textwidth]{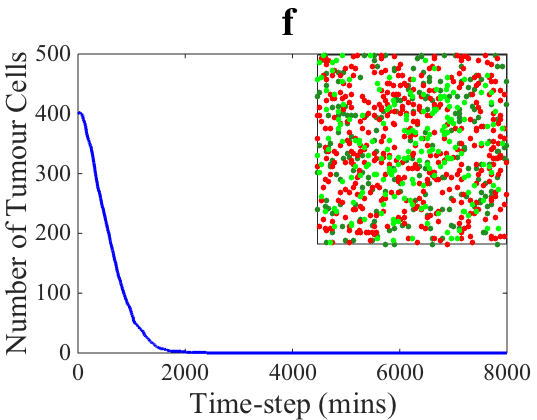}\\
 \includegraphics[width=1\textwidth]{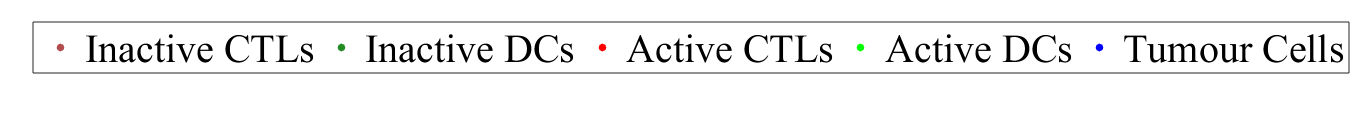} \\\includegraphics[width=1\textwidth]{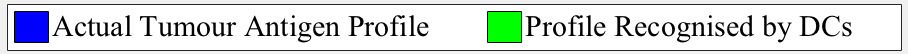}
\caption{\textbf{Increasing $V_{D}$ can result in immune escape or chronic dormancy or immune clearance of the tumour.} Plots in panels \textbf{{{ a-c}}} display the average antigen profile of tumour cells and the average antigen profile recognised by the DCs at the end of simulations. The error lines represent the standard deviation between 5 runs of the simulations. Plots in panels \textbf{{{d-f}}} display the time evolution of the tumour cell number with an example of the observed cell spatial distributions at the final time-step shown in the insets. Three values for the parameter $V_{D}$ are tested: \textbf{a,d} $V_{D}=0.001$, \textbf{b,e} $V_{D}=0.05$ and \textbf{c,f} $V_{D}=0.1$. Here, $V_{T}=0.01$, $\beta=0.01$, $\theta_{E}=\theta_{M}=0$, and all the other parameter values are reported in Table~\ref{tab:2}.}
\label{fig:4} 
\end{figure}

The results discussed in the previous paragraph illustrate how different cell dynamics can be observed for increasing values of the parameter $V_{D}$. We test this further by using the parameter setting of Figure~\ref{fig:3}\nobreak\hspace{0em}b (\emph{i.e.} $V_{T}=0.01$, $\theta_{E}=0$ and $\beta=0.01$) and comparing the dynamics obtained for three different values of $V_{D}$ (\emph{i.e.} $V_{D}=0.001$, $V_{D}=0.05$ and $V_{D}=0.1$). In Figure~\ref{fig:4}, we compare the average antigen profile of the tumour cells at the end of simulations with the average antigen profile recognised by the DCs, and we show the corresponding time evolution of the number of tumour cells. The insets also display the spatial cell distributions observed at the end of simulations to allow for a clearer understanding of the resulting dynamics. We observe that $V_{D}$ is a bifurcation parameter whereby three distinct situations result from choosing increasing values of $V_{D}$. In particular, Figures~\ref{fig:4}\nobreak\hspace{0em}a,d refer to the case where the value of $V_{D}$ is relatively low (\emph{i.e.} $V_{D}=0.001$), and show that there is very little difference between the average antigen profile of the tumour cells and the average recognised antigen profile at the end of simulations. Moreover, after an initial decrease, the tumour cell number increases steadily over time resulting in a relatively large final number of tumour cells. Furthermore, Figures~\ref{fig:4}\nobreak\hspace{0em}b,e refer to the case where an intermediate value of $V_{D}$ is considered (\emph{i.e.} $V_{D}=0.05$), and show that there a is larger variation between the average antigen profile of the tumour cells and the average recognised antigen profile at the end of simulations. Additionally, after a steep decrease, the tumour cell number remains at a low, almost constant, level for the remainder of the simulation time interval. Finally, Figures~\ref{fig:4}\nobreak\hspace{0em}c,f refer to the case where the value of $V_{D}$ is relatively large (\emph{i.e.} $V_{D}=0.1$), and show that the difference between the average antigen profile of the tumour cells. Moreover, the average recognised antigen profile at the end of simulations is even more varied than in the previous cases and the number of tumour cells decreases steadily over time until eventually the tumour is completely removed.

\paragraph{Increasing the T cell receptor binding affinity can benefit the immune system response to cancer}
\label{{sec:3_5}}
\begin{figure}
\includegraphics[width=0.5\textwidth]{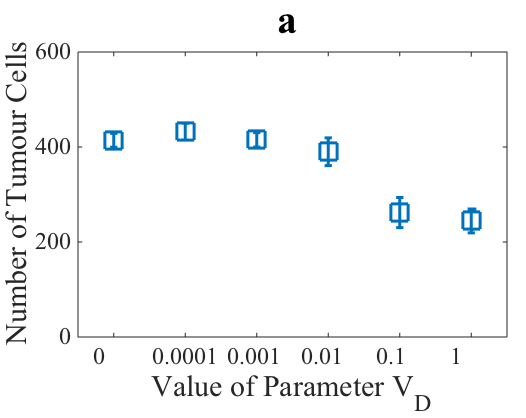}
\includegraphics[width=0.5\textwidth]{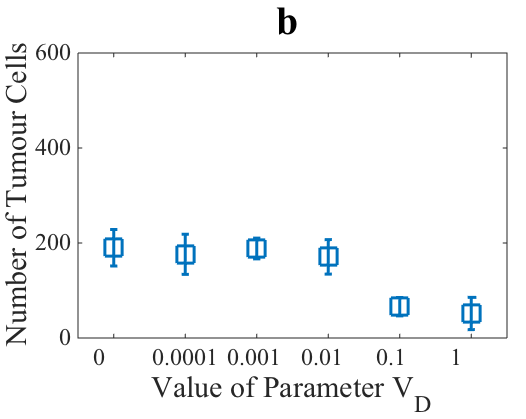}\\
\begin{center}
\includegraphics[width=0.5\textwidth]{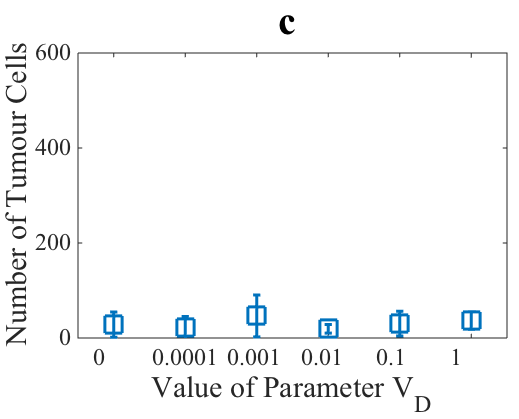}\end{center}
\caption{\textbf{Increasing the T cell receptor binding affinity can benefit the immune system response to cancer.} Plots displaying the number of tumour cells remaining after $1000$ time-steps for increasing values of the parameter $\beta$: \textbf{a} $\beta=0$, \textbf{b} $\beta=0.0001$ and \textbf{c} $\beta=0.001$. For each value of $\beta$ a range of values of the parameter $V_{D}$ are tested. The tumour cell numbers presented have been obtained as the average over $5$ simulations and the error bars display the related standard deviation. Here, $V_T=0.0001$, $\theta_{E}=\theta_{M}=0$, and all the other parameter values are reported in Table~\ref{tab:2}.} 
\label{fig:6} 
\end{figure}

\begin{figure}
\centering
\includegraphics[height=0.32\textwidth]{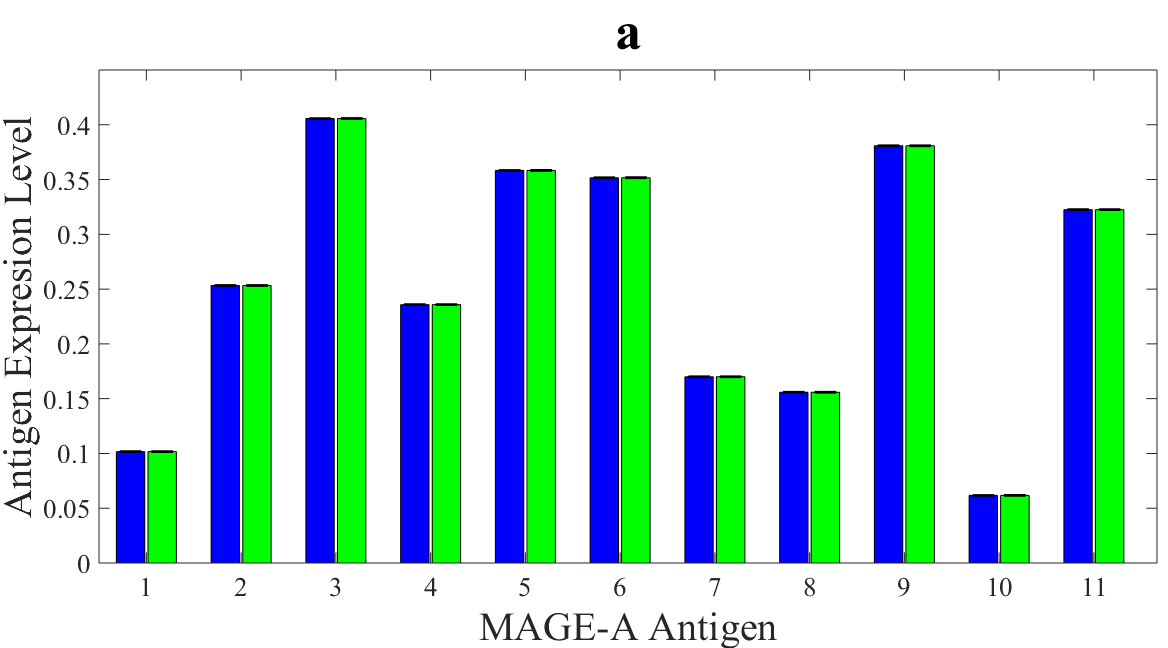}\ \includegraphics[height=0.32\textwidth]{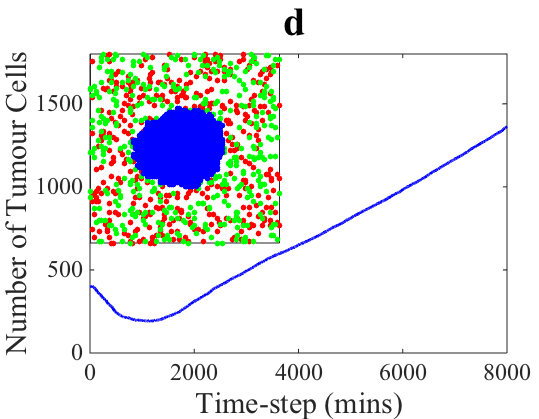}\\
\includegraphics[height=0.32\textwidth]{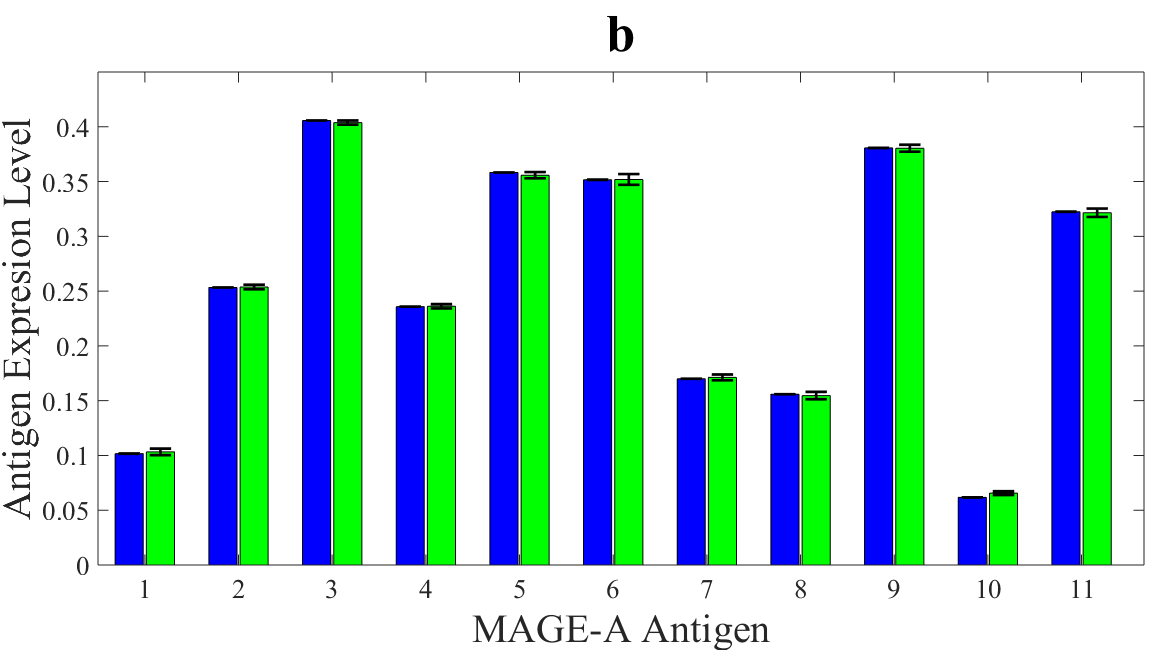}\ \includegraphics[height=0.32\textwidth]{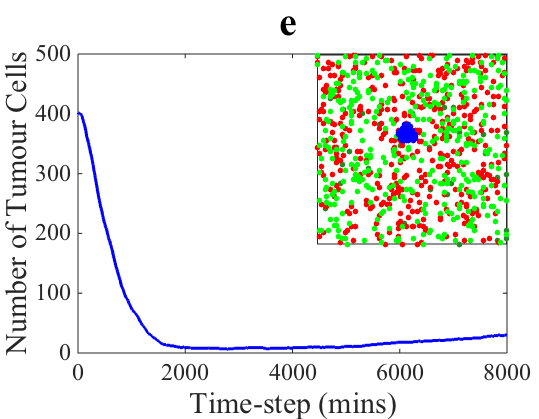}\\
\includegraphics[height=0.32\textwidth]{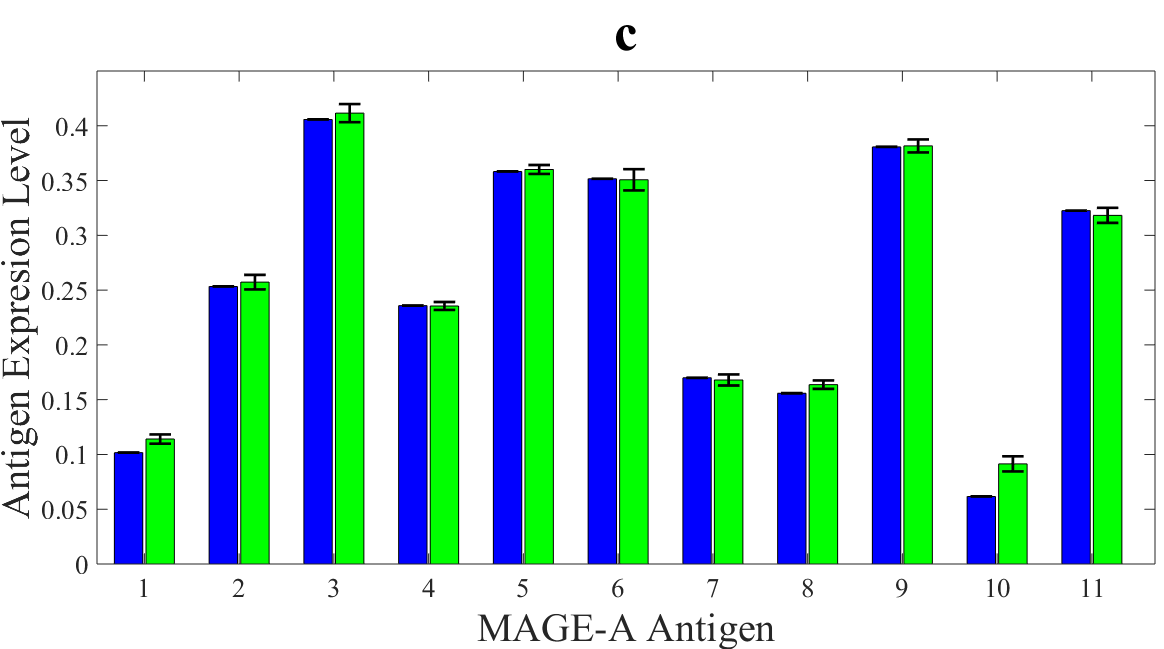}\ \includegraphics[height=0.32\textwidth]{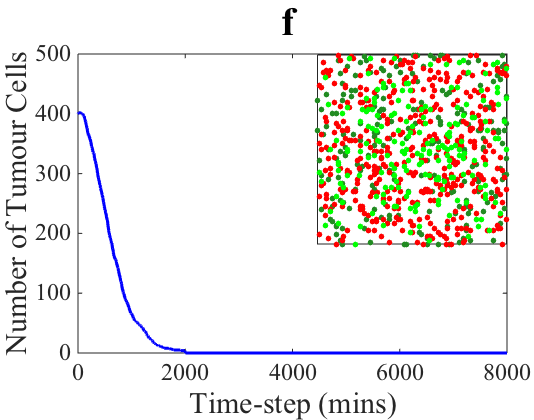}\\
 \includegraphics[width=1\textwidth]{Legend2.png} \\\includegraphics[width=1\textwidth]{Legend1.png}
 \caption{\textbf{{{For multiple parameter settings, increasing $V_{D}$ can result in immune escape or chronic dormancy or immune clearance of the tumour.}}} Plots in panels \textbf{{{ a-c}}} display the average antigen profile of tumour cells and the average antigen profile recognised by the DCs at the end of simulations. The error lines represent the standard deviation between 5 runs of the simulations. Plots in panels \textbf{{{d-f}}} display the time evolution of the tumour cell number with an example of the observed cell spatial distributions at the final time-step shown in the insets. Three values for the parameter $V_{D}$ are tested: \textbf{a,d} $V_{D}=0.001$, \textbf{b,e} $V_{D}=0.05$ and \textbf{c,f} $V_{D}=0.1$. Here, $V_{T}=0.0001$, $\beta=0.0001$, $\theta_{E}=\theta_{M}=0$, and all the other parameter values are reported in Table~\ref{tab:2}.}
\label{fig:7} 
\end{figure}

To explore the effect of altering the T cell receptor binding affinity, we test for three increasing values of the parameter $\beta$ (\emph{i.e.} $\beta=0$, $\beta=0.0001$ and $\beta=0.001$). We choose $V_{T}=0.0001$ and we let the tumour cell antigen profiles remain constant over time (\emph{i.e.} we choose $\theta_{E}=\theta_{M}=0$). For each value of $\beta$ considered, we also explore the effect of increasing the value of the parameter $V_{D}$. In all cases, we carry out numerical simulations for $1000$ time-steps. As shown by Figure~\ref{fig:6}\nobreak\hspace{0em}a, for $\beta=0$, a considerable number of tumour cells remain inside the system at the end of simulations for all values of $V_{D}$. Conversely, the results presented in Figure~\ref{fig:6}\nobreak\hspace{0em}c show that when $\beta$ is sufficiently high, very few tumour cells remain in the system after $1000$ time-steps for all values of $V_{D}$. Moreover, as shown by Figure~\ref{fig:6}\nobreak\hspace{0em}b, for intermediate values of $\beta$, there appears to be a correlation between the number of tumour cells remaining after $1000$ time-steps and the parameter $V_{D}$. In particular, larger values of the parameter $V_{D}$ correspond to smaller numbers of the remaining tumour cells after $1000$ time-steps. These results suggest that the T cell receptor binding affinity plays a key role in the immune response to tumour cells. In the same way as Figure~\ref{fig:4}, Figure~\ref{fig:7} shows that, under the parameter choice of the computational simulations related to Figure~\ref{fig:6}\nobreak\hspace{0em}b, increasing the value of the parameter $V_{D}$ leads to immune escape or chronic dormancy or immune clearance of the tumour.

\paragraph{Increasing the probability of epimutations can lead to variations in the immune response to tumour cells}
\label{sec:3_6}
\begin{figure}
\includegraphics[width=1\textwidth]{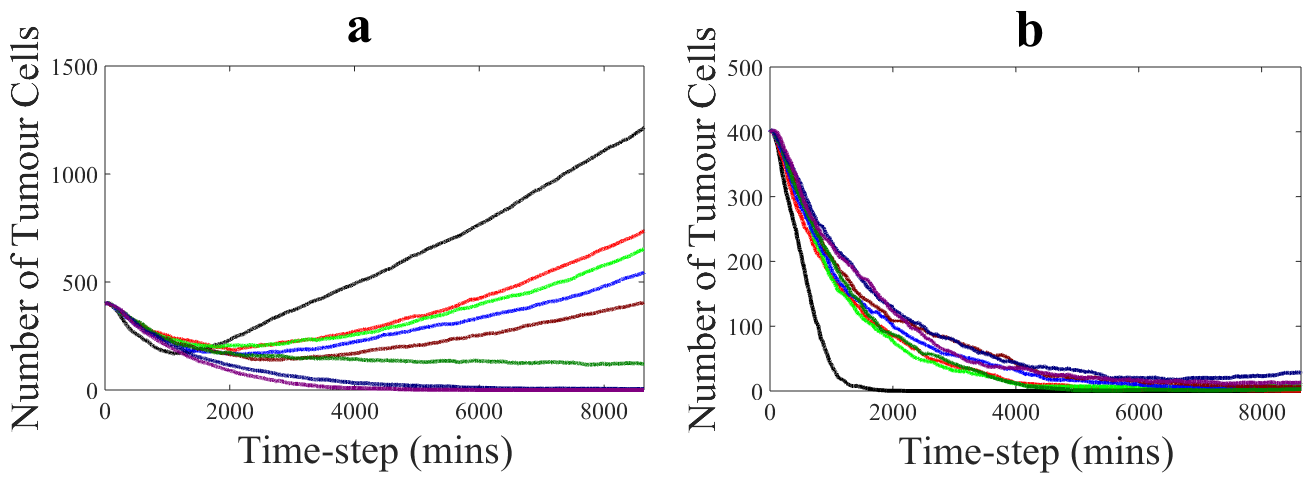}\\
 \includegraphics[width=1\textwidth]{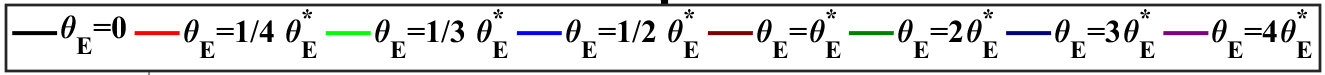}
\caption{\textbf{Increasing the probability of epimutations can lead to variations in the immune response to tumour cells.} Panels \textbf{a} and \textbf{b} display the time evolution of the tumour cell number for increasing values of $\theta_E$ ({\it cf.} the legend below the panels). For the numerical results reported in Panel \textbf{a} all the other parameter values are as for Figures~\ref{fig:4}\nobreak\hspace{0em}a,d and $V_{E}=0.01$, while for the numerical results reported in Panel \textbf{b} all the other parameter values are as for Figures~\ref{fig:7}\nobreak\hspace{0em}c,f and $V_{E}=0.0001$.}
\label{fig:8} 
\end{figure}
So far, we have considered only the situation where the antigen profile of each tumour cell remains constant over time (\emph{i.e.} the probability with which epimutations and mutations leading to antigenic variations occur are $\theta_{E}=0$ and $\theta_{M}=0$). To investigate the effect of epimutations on the success of the immune response against tumour cells, we consider the parameter setting that we have used in the computational simulations shown either in Figures~\ref{fig:4}\nobreak\hspace{0em}a,d or in Figures~\ref{fig:7}\nobreak\hspace{0em}c,f but now we allow the antigen profiles of tumour cells to change through epimutations (\emph{i.e.} we choose $\theta_{E}>0$). We consider eight distinct values of $\theta_{E}$ defined as fractions or multiples of the average probability of epimutations $\theta_{E}^{*}$, given in Table~\ref{tab:2}. {{The range we consider, $0\le\theta_{E}\le0.92$, is chosen to include the range of \cite{DeSmet1996}, $012\le\theta_{E}\le0.45$.} In all cases, we carry out numerical simulations for $8640$ time-steps and we report on tumour cell numbers obtained as the average over $5$ simulations. Figure~\ref{fig:8}a displays the time evolution of the tumour cell number for the parameter setting of Figures~\ref{fig:4}\nobreak\hspace{0em}a,d. These results show that increasing values of $\theta_{E}$ correspond to decreasing numbers of tumour cells inside the system at the end of simulations. In summary, by increasing the probability of epimutations the dynamics of tumour cells change from immune escape, through to chronic dormancy to immune clearance. On the other hand, Figure~\ref{fig:8}b displays the time evolution of the tumour cell number for the parameter setting of Figures~\ref{fig:7}\nobreak\hspace{0em}c,f. These results show that for sufficiently small values of $\theta_{E}$ there are no tumour cells left inside the system at the end of simulations, whereas for larger values of $\theta_{E}$ a small number of tumour cells persist at the final time-step. Generally, by increasing the probability of epimutations the dynamics of tumour cells change from immune clearance to chronic dormancy.

\paragraph{Mutations have a weaker impact on the immune response to tumour cells compared to epimutations}
\label{sec:3_8}
We now compare the impact of mutations and epimutations on the immune response to tumour cells. Following what we have done in the previous paragraph, we consider the parameter setting used in the computational simulations shown either in Figures~\ref{fig:4}\nobreak\hspace{0em}a,d or in Figures~\ref{fig:7}\nobreak\hspace{0em}c,f but now we allow the antigen profiles of tumour cells to change through mutations (\emph{i.e.} we choose $\theta_{M}>0$). We consider eight distinct values of $\theta_{M}$ defined as fractions or multiples of the average probability of epimutations $\theta_{E}^{*}$, given in Table~\ref{tab:2}. In all cases, we carry out numerical simulations for $8640$ time-steps and we report on tumour cell numbers obtained as the average over $5$ simulations. Figure~\ref{fig:10}a displays the time evolution of the tumour cell number for the parameter setting of Figures~\ref{fig:4}\nobreak\hspace{0em}a,d and shows that immune escape occurs for all values of $\theta_M$ considered. On the other hand, Figure~\ref{fig:10}b displays the time evolution of the tumour cell number for the parameter setting of Figures~\ref{fig:7}\nobreak\hspace{0em}c,f and shows that immune clearance occurs for all values of $\theta_M$ considered. Comparing these results with those displayed in Figure~\ref{fig:8}a and Figure~\ref{fig:8}b, respectively, we reach the conclusion that mutations have a weaker impact on the immune response to tumour cells compared to epimutations.
\begin{figure}
\includegraphics[width=1\textwidth]{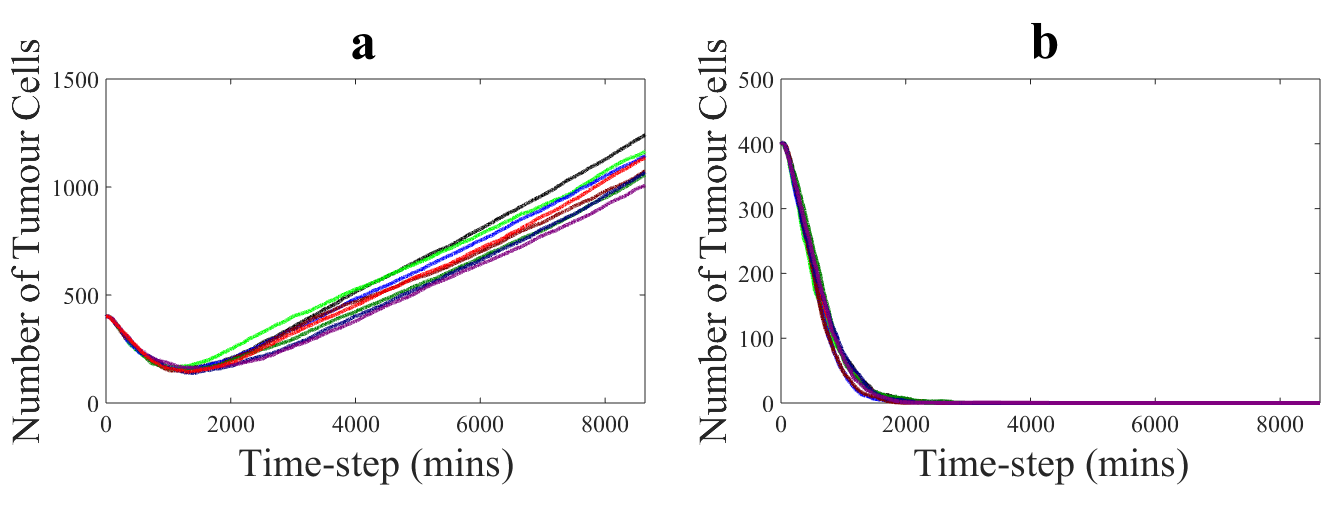}\\
 \includegraphics[width=1\textwidth]{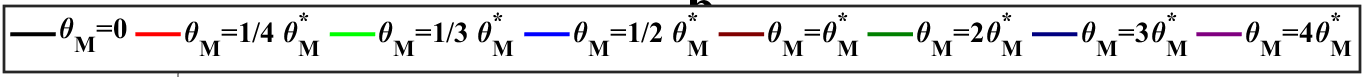}
 \caption{\textbf{Mutations have a weaker impact on the immune response to tumour cells compared to epimutations.} Panels \textbf{a} and \textbf{b} display the time evolution of the tumour cell number for increasing values of $\theta_M$ ({\it cf.} the legend below the panels). Here $\theta^*_M = \theta^*_E$. For the numerical results reported in Panel \textbf{a} all the other parameter values are as for Figures~\ref{fig:4}\nobreak\hspace{0em}a,d and $V_{M}=0.01$, while for the numerical results reported in Panel \textbf{b} all the other parameter values are as for Figures~\ref{fig:7}\nobreak\hspace{0em}c,f and $V_{M}=0.0001$.}
\label{fig:10} 
\end{figure}

\section{Discussion and conclusions}
\label{sec:4}
Spatial interactions between cancer and immune cells, as well as the recognition of tumour antigens by cells of the immune system, play a key role in the immune response against solid tumours. The existing mathematical models generally focus only on one of these key aspects. We have presented here a spatially explicit stochastic individual-based model that incorporates the adaptive processes driving tumour antigen recognition. Our model takes explicitly into account the dynamical heterogeneity of tumour antigen expression, and effectively captures the way in which this affects the immune response against the tumour. 

Our computational simulation results show that the initial antigen expression profiles of cancer cells within the tumour have a crucial impact upon the outcome of the immune response {{[refer to Figure~\ref{fig:3}]}}. In the situation of an almost homogeneous tumour (\emph{i.e.} where all tumour cells have a similar antigen profile), immune clearance occurs. Conversely, when the antigenic composition between cancer cells is highly heterogeneous the tumour may be able to escape the immune system response and continue growing. Interestingly, for moderate levels of initial antigenic heterogeneity our results demonstrate that the fate of the tumour is determined by the specificity of the cellular immune response. 

The computational outcomes of our model indicate that the parameter controlling the specificity of the antigen recognition process of the dendritic cells (\emph{i.e.} the parameter $V_D$) ultimately dictates which receptors are produced by the cytotoxic T lymphocytes {{[refer to Figures~\ref{fig:3}~and~\ref{fig:4}]}}. A larger value of this parameter brings about a more diverse receptor repertoire of the CTLs, and in turn results in a better immune response. This suggests that it is advantageous for the T cell pool to be multi-specific, whereby several different antigen receptors are simultaneously expressed by the CTL pool. In this respect, the outcomes of our model recapitulate the conclusions of experimental papers showing the success of a more diverse T cell repertoire in response to cancer~\citep{Carreno2015,Gerdemann2011,Ott2017,Sahin2017,Schumacher2016,Sharma2015}. {{We remark that new experimental techniques have recently been developed to alter the specificity of T cell receptors~\citep{smith2014}. One particular approach is to use gene editing, \emph{in vitro}, to modify which antigen the T cell receptors will respond to~\citep{albers2019}.}}

Moreover, our numerical results support the idea that varying the specificity of the immune response can result in three distinct scenarios, from immune escape, through to chronic dormancy to immune clearance of the tumour {{[refer to Figures~\ref{fig:4}~and~\ref{fig:7}]}}. The importance of tumour dormancy controlled by the immune system (\emph{i.e.} immunological dormancy) has been highlighted by previous experimental and theoretical work~\citep{Kuznetsov1994,Lorenzi2015,Matzavinos2004,Wu2018}. In particular, immunological dormancy can explain situations where there is an extended period of time before the occurrence of tumour relapse~\citep{Aguirre2007,Gomis2017,Manjili2018,Teng2008,Wang2013,Yeh2015}. In this regard, our model suggests the existence of a possible relationship between the specificity of the immune response and the emergence of prolonged immunological dormancy.
 
We have also explored the way in which altering the binding affinity of the CTLs to their corresponding tumour antigen may change the immune response to the tumour. Our results indicate that a stronger binding affinity leads to a more effective immune response, as the CTLs have a wider range of tumour cells that they can interact with {{[refer to Figure~\ref{fig:6}]}}. Previously, \cite{Gerdemann2011} found experimentally that a strong T cell binding affinity to tumour antigens played a key role in the overall immune response to the disease. Integrating the outcomes of our model and such experimental findings suggests that enhancing the binding affinity of T cells -- \emph{e.g.} through the modification of the receptors that the T cells of a patient can produce -- could be a potential target of adoptive T cell therapy.

The results from our computational simulations suggest that changes in the antigenic expression of tumour cells due to epimutations can be either beneficial or detrimental to the immune response to a solid tumour {{[refer to Figure~\ref{fig:8}]}}. In more detail, we have found that in some cases increasing the probability of epimutations could transform situations of immune escape into tumour dormancy and eventually tumour removal. These findings are interesting in light of cancer therapy as they suggest that the efficacy of the immune response against solid tumours could be enhanced by increasing the probability of epimutations. In this respect, the loss of DNA methylation was the first epimutation to be identified in cancer cells \citep{Feinberg2004b} and several experimental and clinical works found that the expression of MAGE antigens could be increased through demethylation~\citep{Chinnasamy2011,Gerdemann2011,Graff2002,Wischnewski2006}. Taken together, the outcomes of our model suggest that by combining a T cell therapy targeting multiple MAGE genes -- \emph{e.g.} using approaches similar to those of \cite{Gerdemann2011} -- and increasing the probability of epimutations through demethylating agents -- \emph{e.g.} using methods similar to those of \cite{Chinnasamy2011} and \cite{Wischnewski2006} -- a stronger immune response could be induced. However, in other cases, we have observed that increasing the probability of epimutations can turn instances of tumour removal into scenarios whereby a small number of tumour cells persisted over time. These contrasting results were also suggested previously through experimental research, where epimutations could be either beneficial or detrimental to tumour development~\citep{Chen2017,Yarchoan2017}. 

We have additionally studied the effect of variations in the antigenic expression of tumour cells caused by mutations. In all parameter settings we have considered, increasing the probability of mutations did not change the resulting dynamics of the tumour-immune response {{[refer to Figure~\ref{fig:10}]}}. This suggests that mutations have a weaker effect on tumour-immune competition than antigenic variations caused by epimutations. This finding is coherent with experimental observations indicating that epimutations generally occur more frequently than mutations in tumour development \citep{Feinberg2004a,Peltomaki2012}. Hence, our results demonstrate the importance of understanding the underlying causes of antigenic variations in tumour cells when considering tumour-immune competition. 

Looking to the future, our individual-based model could be developed further in several ways. We could incorporate extended aspects of the tumour micro-environment, such as, proliferation of the immune cells and the interaction of tumour cells with abiotic factors (\emph{e.g.} oxygen and glucose) that can affect the phenotypic composition of the tumour. Moreover, by posing the model on a 3D domain, further understanding could be obtained regarding the spatial dynamics of the tumour-immune response. The flexibility of our model would also allow for the inclusion of a wider range of antigens. Finally, although useful for investigating the qualitative and quantitative dynamics of a biological system, individual-based models like ours are not amenable to mathematical analysis. In a variety of contexts the benefit of relating stochastic individual-based models to deterministic continuum models has been highlighted \citep{Champagnat2006,Chisholm2016,Chisholm2015,Deroulers2009,Painter2015,Penington2011,Stevens2000}. Combining such modelling approaches makes it possible to integrate computational simulations with analytical results, thus enabling a more extensive exploration of the model parameter space. This is a line of research that we will be pursuing in the near future by exploiting the formal methods that we have recently presented~\citep{macfarlane2018bridging,stace2019}.

\section*{Acknowledgments}
FRM is funded by the Engineering and Physical Sciences Research Council (EPSRC). 

\section*{Bibliography}
\label{sec:5}
\bibliography{Manuscript_Revision}

\begin{thebibliography}{105}
\expandafter\ifx\csname natexlab\endcsname\relax\def\natexlab#1{#1}\fi
\providecommand{\url}[1]{\texttt{#1}}
\providecommand{\href}[2]{#2}
\providecommand{\path}[1]{#1}
\providecommand{\DOIprefix}{doi:}
\providecommand{\ArXivprefix}{arXiv:}
\providecommand{\URLprefix}{URL: }
\providecommand{\Pubmedprefix}{pmid:}
\providecommand{\doi}[1]{\href{http://dx.doi.org/#1}{\path{#1}}}
\providecommand{\Pubmed}[1]{\href{pmid:#1}{\path{#1}}}
\providecommand{\bibinfo}[2]{#2}
\ifx\xfnm\relax \def\xfnm[#1]{\unskip,\space#1}\fi
\bibitem[{Aguirre-Ghiso(2007)}]{Aguirre2007}
\bibinfo{author}{Aguirre-Ghiso, J.A.}, \bibinfo{year}{2007}.
\newblock \bibinfo{title}{Models, mechanisms and clinical evidence for cancer
  dormancy}.
\newblock \bibinfo{journal}{Nat Rev Cancer} \bibinfo{volume}{7},
  \bibinfo{pages}{834--846}.
\bibitem[{Al-Tameemi et~al.(2012)Al-Tameemi, Chaplain and
  d'Onofrio}]{Al-Tameemi2012}
\bibinfo{author}{Al-Tameemi, M.}, \bibinfo{author}{Chaplain, M.A.J.},
  \bibinfo{author}{d'Onofrio, A.}, \bibinfo{year}{2012}.
\newblock \bibinfo{title}{Evasion of tumours from the control of the immune
  system: \uppercase{C}onsequences of brief encounters}.
\newblock \bibinfo{journal}{Biol Direct} \bibinfo{volume}{7},
  \bibinfo{pages}{31}.
\bibitem[{Albers et~al.(2019)Albers, Ammon, Gosmann, Audehm, Thoene, Winter,
  Secci, Wolf, Stelzl, Steiger et~al.}]{albers2019}
\bibinfo{author}{Albers, J.J.}, \bibinfo{author}{Ammon, T.},
  \bibinfo{author}{Gosmann, D.}, \bibinfo{author}{Audehm, S.},
  \bibinfo{author}{Thoene, S.}, \bibinfo{author}{Winter, C.},
  \bibinfo{author}{Secci, R.}, \bibinfo{author}{Wolf, A.},
  \bibinfo{author}{Stelzl, A.}, \bibinfo{author}{Steiger, K.}, et~al.,
  \bibinfo{year}{2019}.
\newblock \bibinfo{title}{Gene editing enables \uppercase{T}-cell engineering
  to redirect antigen specificity for potent tumor rejection}.
\newblock \bibinfo{journal}{Life Sci Alliance} \bibinfo{volume}{2},
  \bibinfo{pages}{e201900367}.
\bibitem[{Arciero et~al.(2004)Arciero, Jackson and Kirschner}]{Arciero2004}
\bibinfo{author}{Arciero, J.C.}, \bibinfo{author}{Jackson, T.L.},
  \bibinfo{author}{Kirschner, D.E.}, \bibinfo{year}{2004}.
\newblock \bibinfo{title}{A mathematical model of tumor-immune evasion and
  si\uppercase{RNA} treatment}.
\newblock \bibinfo{journal}{Discrete Contin Dyn Syst Ser B}
  \bibinfo{volume}{4}, \bibinfo{pages}{39--58}.
\bibitem[{Asatryan and Komarova(2016)}]{Asatryan2016}
\bibinfo{author}{Asatryan, A.D.}, \bibinfo{author}{Komarova, N.L.},
  \bibinfo{year}{2016}.
\newblock \bibinfo{title}{Evolution of genetic instability in heterogeneous
  tumors}.
\newblock \bibinfo{journal}{J Theor Biol} \bibinfo{volume}{396},
  \bibinfo{pages}{1--12}.
\bibitem[{Balachandran et~al.(2017)Balachandran, {\L}uksza, Zhao, Makarov,
  Moral, Remark et~al.}]{Balachandran2017}
\bibinfo{author}{Balachandran, V.P.}, \bibinfo{author}{{\L}uksza, M.},
  \bibinfo{author}{Zhao, J.N.}, \bibinfo{author}{Makarov, V.},
  \bibinfo{author}{Moral, J.A.}, \bibinfo{author}{Remark, R.}, et~al.,
  \bibinfo{year}{2017}.
\newblock \bibinfo{title}{Identification of unique neoantigen qualities in
  long-term survivors of pancreatic cancer}.
\newblock \bibinfo{journal}{Nature} \bibinfo{volume}{551},
  \bibinfo{pages}{512--516}.
\bibitem[{Balea et~al.(2014)Balea, Halanay, Jardan, Neam{\c{t}}u and
  Safta}]{Balea2014}
\bibinfo{author}{Balea, S.}, \bibinfo{author}{Halanay, A.},
  \bibinfo{author}{Jardan, D.}, \bibinfo{author}{Neam{\c{t}}u, M.},
  \bibinfo{author}{Safta, C.}, \bibinfo{year}{2014}.
\newblock \bibinfo{title}{Stability analysis of a feedback model for the action
  of the immune system in leukemia}.
\newblock \bibinfo{journal}{Math Model Nat Phenom} \bibinfo{volume}{9},
  \bibinfo{pages}{108--132}.
\bibitem[{Besse et~al.(2018)Besse, Clapp, Bernard, Nicolini, Levy and
  Lepoutre}]{Besse2018}
\bibinfo{author}{Besse, A.}, \bibinfo{author}{Clapp, G.D.},
  \bibinfo{author}{Bernard, S.}, \bibinfo{author}{Nicolini, F.E.},
  \bibinfo{author}{Levy, D.}, \bibinfo{author}{Lepoutre, T.},
  \bibinfo{year}{2018}.
\newblock \bibinfo{title}{Stability analysis of a model of interaction between
  the immune system and cancer cells in chronic myelogenous leukemia}.
\newblock \bibinfo{journal}{Bull Math Biol} \bibinfo{volume}{80},
  \bibinfo{pages}{1084--1110}.
\bibitem[{Bianca et~al.(2012)Bianca, Chiacchio, Pappalardo and
  Pennisi}]{Bianca2012}
\bibinfo{author}{Bianca, C.}, \bibinfo{author}{Chiacchio, F.},
  \bibinfo{author}{Pappalardo, F.}, \bibinfo{author}{Pennisi, M.},
  \bibinfo{year}{2012}.
\newblock \bibinfo{title}{Mathematical modeling of the immune system
  recognition to mammary carcinoma antigen}.
\newblock \bibinfo{journal}{BMC Bioinformatics} \bibinfo{volume}{13},
  \bibinfo{pages}{S21}.
\bibitem[{Boes et~al.(2002)Boes, Cerny, Massol, den Brouw, Kirchhausen, Chen
  and Ploegh}]{boes2002}
\bibinfo{author}{Boes, M.}, \bibinfo{author}{Cerny, J.},
  \bibinfo{author}{Massol, R.}, \bibinfo{author}{den Brouw, M.O.P.},
  \bibinfo{author}{Kirchhausen, T.}, \bibinfo{author}{Chen, J.},
  \bibinfo{author}{Ploegh, H.L.}, \bibinfo{year}{2002}.
\newblock \bibinfo{title}{T-cell engagement of dendritic cells rapidly
  rearranges \uppercase{MHC} class \uppercase{II} transport}.
\newblock \bibinfo{journal}{Nature} \bibinfo{volume}{418},
  \bibinfo{pages}{983}.
\bibitem[{Boissonnas et~al.(2007)Boissonnas, Fetler, Zeelenberg, Hugues and
  Amigorena}]{Boissonnas2007}
\bibinfo{author}{Boissonnas, A.}, \bibinfo{author}{Fetler, L.},
  \bibinfo{author}{Zeelenberg, I.S.}, \bibinfo{author}{Hugues, S.},
  \bibinfo{author}{Amigorena, S.}, \bibinfo{year}{2007}.
\newblock \bibinfo{title}{In vivo imaging of cytotoxic \uppercase{T} cell
  infiltration and elimination of a solid tumor}.
\newblock \bibinfo{journal}{J Exp Med} \bibinfo{volume}{204},
  \bibinfo{pages}{345--356}.
\bibitem[{Boon et~al.(2006)Boon, Coulie, Van~den Eynde and van~der
  Bruggen}]{Boon2006}
\bibinfo{author}{Boon, T.}, \bibinfo{author}{Coulie, P.G.},
  \bibinfo{author}{Van~den Eynde, B.J.}, \bibinfo{author}{van~der Bruggen, P.},
  \bibinfo{year}{2006}.
\newblock \bibinfo{title}{Human \uppercase{T} cell responses against melanoma}.
\newblock \bibinfo{journal}{Annu Rev Immunol} \bibinfo{volume}{24},
  \bibinfo{pages}{175--208}.
\bibitem[{Bouchnita et~al.(2017)Bouchnita, Belmaati, Aboulaich, Koury and
  Volpert}]{Bouchnita2017}
\bibinfo{author}{Bouchnita, A.}, \bibinfo{author}{Belmaati, F.E.},
  \bibinfo{author}{Aboulaich, R.}, \bibinfo{author}{Koury, M.J.},
  \bibinfo{author}{Volpert, V.}, \bibinfo{year}{2017}.
\newblock \bibinfo{title}{A hybrid computation model to describe the
  progression of multiple myeloma and its intra-clonal heterogeneity}.
\newblock \bibinfo{journal}{Computation} \bibinfo{volume}{5},
  \bibinfo{pages}{16}.
\bibitem[{Brenner et~al.(2008)Brenner, Krammer and Arnold}]{Brenner2008}
\bibinfo{author}{Brenner, D.}, \bibinfo{author}{Krammer, P.H.},
  \bibinfo{author}{Arnold, R.}, \bibinfo{year}{2008}.
\newblock \bibinfo{title}{Concepts of activated \uppercase{T} cell death}.
\newblock \bibinfo{journal}{Crit Rev Oncol Hematol} \bibinfo{volume}{66},
  \bibinfo{pages}{52--64}.
\bibitem[{Carreno et~al.(2015)Carreno, Magrini, Becker-Hapak, Kaabinejadian,
  Hundal, Petti, Ly, Lie, Hildebrand, Mardis et~al.}]{Carreno2015}
\bibinfo{author}{Carreno, B.M.}, \bibinfo{author}{Magrini, V.},
  \bibinfo{author}{Becker-Hapak, M.}, \bibinfo{author}{Kaabinejadian, S.},
  \bibinfo{author}{Hundal, J.}, \bibinfo{author}{Petti, A.A.},
  \bibinfo{author}{Ly, A.}, \bibinfo{author}{Lie, W.R.},
  \bibinfo{author}{Hildebrand, W.H.}, \bibinfo{author}{Mardis, E.R.}, et~al.,
  \bibinfo{year}{2015}.
\newblock \bibinfo{title}{A dendritic cell vaccine increases the breadth and
  diversity of melanoma neoantigen-specific \uppercase{T} cells}.
\newblock \bibinfo{journal}{Science} \bibinfo{volume}{348},
  \bibinfo{pages}{803--808}.
\bibitem[{Chalitchagorn et~al.(2004)Chalitchagorn, Shuangshoti, Hourpai,
  Kongruttanachok, Tangkijvanich, Thong-ngam, Voravud, Sriuranpong and
  Mutirangura}]{Chalitchagorn2004}
\bibinfo{author}{Chalitchagorn, K.}, \bibinfo{author}{Shuangshoti, S.},
  \bibinfo{author}{Hourpai, N.}, \bibinfo{author}{Kongruttanachok, N.},
  \bibinfo{author}{Tangkijvanich, P.}, \bibinfo{author}{Thong-ngam, D.},
  \bibinfo{author}{Voravud, N.}, \bibinfo{author}{Sriuranpong, V.},
  \bibinfo{author}{Mutirangura, A.}, \bibinfo{year}{2004}.
\newblock \bibinfo{title}{Distinctive pattern of\uppercase{ LINE-1} methylation
  level in normal tissues and the association with carcinogenesis}.
\newblock \bibinfo{journal}{Oncogene} \bibinfo{volume}{23},
  \bibinfo{pages}{8841}.
\bibitem[{Champagnat et~al.(2006)Champagnat, Ferri{\`e}re and
  M{\'e}l{\'e}ard}]{Champagnat2006}
\bibinfo{author}{Champagnat, N.}, \bibinfo{author}{Ferri{\`e}re, R.},
  \bibinfo{author}{M{\'e}l{\'e}ard, S.}, \bibinfo{year}{2006}.
\newblock \bibinfo{title}{Unifying evolutionary dynamics: from individual
  stochastic processes to macroscopic models}.
\newblock \bibinfo{journal}{Theor Pop Biol} \bibinfo{volume}{69},
  \bibinfo{pages}{297--321}.
\bibitem[{Chaplain et~al.(2018)Chaplain, Lorenzi and
  Macfarlane}]{macfarlane2018bridging}
\bibinfo{author}{Chaplain, M.A.}, \bibinfo{author}{Lorenzi, T.},
  \bibinfo{author}{Macfarlane, F.R.}, \bibinfo{year}{2018}.
\newblock \bibinfo{title}{Bridging the gap between individual-based and
  continuum models of growing cell populations}.
\newblock \bibinfo{journal}{arXiv preprint arXiv:1812.05872} .
\bibitem[{Chaplin(2010)}]{Chaplin2010}
\bibinfo{author}{Chaplin, D.D.}, \bibinfo{year}{2010}.
\newblock \bibinfo{title}{Overview of the immune response}.
\newblock \bibinfo{journal}{J Allergy Clin Immunol} \bibinfo{volume}{2},
  \bibinfo{pages}{S3--S23}.
\bibitem[{Chen and Mellman(2017)}]{Chen2017}
\bibinfo{author}{Chen, D.S.}, \bibinfo{author}{Mellman, I.},
  \bibinfo{year}{2017}.
\newblock \bibinfo{title}{Elements of cancer immunity and the cancer-immune set
  point}.
\newblock \bibinfo{journal}{Nature} \bibinfo{volume}{541},
  \bibinfo{pages}{321--330}.
\bibitem[{Chinnasamy et~al.(2011)Chinnasamy, Wargo, Yu, Rao, Frankel, Riley,
  Hong, Parkhurst, Feldman, Schrump et~al.}]{Chinnasamy2011}
\bibinfo{author}{Chinnasamy, N.}, \bibinfo{author}{Wargo, J.A.},
  \bibinfo{author}{Yu, Z.}, \bibinfo{author}{Rao, M.},
  \bibinfo{author}{Frankel, T.L.}, \bibinfo{author}{Riley, J.P.},
  \bibinfo{author}{Hong, J.J.}, \bibinfo{author}{Parkhurst, M.R.},
  \bibinfo{author}{Feldman, S.A.}, \bibinfo{author}{Schrump, D.S.}, et~al.,
  \bibinfo{year}{2011}.
\newblock \bibinfo{title}{A \uppercase{TCR} targeting the
  \uppercase{HLA-A}$^{*}$ 0201--restricted epitope of \uppercase{MAGE-A3}
  recognizes multiple epitopes of the \uppercase{MAGE-A} antigen superfamily in
  several types of cancer}.
\newblock \bibinfo{journal}{J Immunol} \bibinfo{volume}{186},
  \bibinfo{pages}{685--696}.
\bibitem[{Chisholm et~al.(2016)Chisholm, Lorenzi, Desvillettes and
  Hughes}]{Chisholm2016}
\bibinfo{author}{Chisholm, R.H.}, \bibinfo{author}{Lorenzi, T.},
  \bibinfo{author}{Desvillettes, L.}, \bibinfo{author}{Hughes, B.D.},
  \bibinfo{year}{2016}.
\newblock \bibinfo{title}{Evolutionary dynamics of phenotype-structured
  populations: From individual-level mechanisms to population-level
  consequences}.
\newblock \bibinfo{journal}{Z Angew Math Phys} \bibinfo{volume}{67},
  \bibinfo{pages}{100}.
\bibitem[{Chisholm et~al.(2015)Chisholm, Lorenzi, Lorz, Larsen, Almeida,
  Escargueil and Clairambault}]{Chisholm2015}
\bibinfo{author}{Chisholm, R.H.}, \bibinfo{author}{Lorenzi, T.},
  \bibinfo{author}{Lorz, A.}, \bibinfo{author}{Larsen, A.K.},
  \bibinfo{author}{Almeida, L.}, \bibinfo{author}{Escargueil, A.},
  \bibinfo{author}{Clairambault, J.}, \bibinfo{year}{2015}.
\newblock \bibinfo{title}{Emergence of drug tolerance in cancer cell
  populations: An evolutionary outcome of selection, non-genetic instability
  and stress-induced adaptation}.
\newblock \bibinfo{journal}{Cancer Res} \bibinfo{volume}{75},
  \bibinfo{pages}{930--939}.
\bibitem[{Christophe et~al.(2015)Christophe, M{\"u}ller, Rodrigues, Petit,
  Cattiaux, Dupr{\'e}, Gadat and Valitutti}]{Christophe2015}
\bibinfo{author}{Christophe, C.}, \bibinfo{author}{M{\"u}ller, S.},
  \bibinfo{author}{Rodrigues, M.}, \bibinfo{author}{Petit, A.E.},
  \bibinfo{author}{Cattiaux, P.}, \bibinfo{author}{Dupr{\'e}, L.},
  \bibinfo{author}{Gadat, S.}, \bibinfo{author}{Valitutti, S.},
  \bibinfo{year}{2015}.
\newblock \bibinfo{title}{A biased competition theory of cytotoxic
  \uppercase{T} lymphocyte interaction with tumor nodules}.
\newblock \bibinfo{journal}{PloS One} \bibinfo{volume}{10},
  \bibinfo{pages}{e0120053}.
\bibitem[{Coico and Sunshine(2015)}]{Coico2015a}
\bibinfo{author}{Coico, R.}, \bibinfo{author}{Sunshine, G.},
  \bibinfo{year}{2015}.
\newblock \bibinfo{title}{Overview of the immune system}, in:
  \bibinfo{booktitle}{Immunology: A short course 7th Edition}.
  \bibinfo{publisher}{John Wiley \& Sons}. chapter~\bibinfo{chapter}{1}, pp.
  \bibinfo{pages}{1--11}.
\bibitem[{Connerotte et~al.(2008)Connerotte, Van~Pel, Godelaine, Tartour,
  Schuler-Thurner, Lucas, Thielemans, Schuler and Coulie}]{Connerotte2008}
\bibinfo{author}{Connerotte, T.}, \bibinfo{author}{Van~Pel, A.},
  \bibinfo{author}{Godelaine, D.}, \bibinfo{author}{Tartour, E.},
  \bibinfo{author}{Schuler-Thurner, B.}, \bibinfo{author}{Lucas, S.},
  \bibinfo{author}{Thielemans, K.}, \bibinfo{author}{Schuler, G.},
  \bibinfo{author}{Coulie, P.G.}, \bibinfo{year}{2008}.
\newblock \bibinfo{title}{Functions of anti-\uppercase{MAGE T}-cells induced in
  melanoma patients under different vaccination modalities}.
\newblock \bibinfo{journal}{Cancer Res} \bibinfo{volume}{68},
  \bibinfo{pages}{3931--3940}.
\bibitem[{Costa et~al.(2007)Costa, Le~Blanc and Brodin}]{costa2007}
\bibinfo{author}{Costa, F.F.}, \bibinfo{author}{Le~Blanc, K.},
  \bibinfo{author}{Brodin, B.}, \bibinfo{year}{2007}.
\newblock \bibinfo{title}{Concise review: \uppercase{C}ancer/testis antigens,
  stem cells, and cancer}.
\newblock \bibinfo{journal}{Stem Cells} \bibinfo{volume}{25},
  \bibinfo{pages}{707--711}.
\bibitem[{Coulie et~al.(2014)Coulie, Van~den Eynde, Van Der~Bruggen and
  Boon}]{Coulie2014}
\bibinfo{author}{Coulie, P.G.}, \bibinfo{author}{Van~den Eynde, B.J.},
  \bibinfo{author}{Van Der~Bruggen, P.}, \bibinfo{author}{Boon, T.},
  \bibinfo{year}{2014}.
\newblock \bibinfo{title}{Tumour antigens recognized by \uppercase{T}
  lymphocytes: \uppercase{A}t the core of cancer immunotherapy}.
\newblock \bibinfo{journal}{Nat Rev Cancer} \bibinfo{volume}{14},
  \bibinfo{pages}{135}.
\bibitem[{Couzin-Frankel(2013)}]{frankel2013}
\bibinfo{author}{Couzin-Frankel, J.}, \bibinfo{year}{2013}.
\newblock \bibinfo{title}{Cancer immunotherapy}.
\newblock \bibinfo{journal}{Science} \bibinfo{volume}{342},
  \bibinfo{pages}{1432--1433}.
\bibitem[{Daudi et~al.(2014)Daudi, Eng, Mhawech-Fauceglia, Morrison, Miliotto,
  Beck, Matsuzaki, Tsuji, Groman, Gnjatic et~al.}]{daudi2014}
\bibinfo{author}{Daudi, S.}, \bibinfo{author}{Eng, K.H.},
  \bibinfo{author}{Mhawech-Fauceglia, P.}, \bibinfo{author}{Morrison, C.},
  \bibinfo{author}{Miliotto, A.}, \bibinfo{author}{Beck, A.},
  \bibinfo{author}{Matsuzaki, J.}, \bibinfo{author}{Tsuji, T.},
  \bibinfo{author}{Groman, A.}, \bibinfo{author}{Gnjatic, S.}, et~al.,
  \bibinfo{year}{2014}.
\newblock \bibinfo{title}{Expression and immune responses to \uppercase{MAGE}
  antigens predict survival in epithelial ovarian cancer}.
\newblock \bibinfo{journal}{PloS One} \bibinfo{volume}{9},
  \bibinfo{pages}{e104099}.
\bibitem[{Davis et~al.(1998)Davis, Boniface, Reich, Lyons, Hampl, Arden and
  Chien}]{davis1998}
\bibinfo{author}{Davis, M.M.}, \bibinfo{author}{Boniface, J.J.},
  \bibinfo{author}{Reich, Z.}, \bibinfo{author}{Lyons, D.},
  \bibinfo{author}{Hampl, J.}, \bibinfo{author}{Arden, B.},
  \bibinfo{author}{Chien, Y.}, \bibinfo{year}{1998}.
\newblock \bibinfo{title}{Ligand recognition by $\alpha$$\beta$ \uppercase{T}
  cell receptors}.
\newblock \bibinfo{journal}{Annu Rev Immunol} \bibinfo{volume}{16},
  \bibinfo{pages}{523--544}.
\bibitem[{De~Boer et~al.(1985)De~Boer, Hogeweg, Dullens, De~Weger and
  Den~Otter}]{DeBoer1985}
\bibinfo{author}{De~Boer, R.J.}, \bibinfo{author}{Hogeweg, P.},
  \bibinfo{author}{Dullens, H.F.}, \bibinfo{author}{De~Weger, R.A.},
  \bibinfo{author}{Den~Otter, W.}, \bibinfo{year}{1985}.
\newblock \bibinfo{title}{Macrophage \uppercase{T} lymphocyte interactions in
  the anti-tumor immune response: \uppercase{A} mathematical model.}
\newblock \bibinfo{journal}{J Immunol} \bibinfo{volume}{134},
  \bibinfo{pages}{2748--2758}.
\bibitem[{De~Smet et~al.(1996)De~Smet, De~Backer, Faraoni, Lurquin, Brasseur
  and Boon}]{DeSmet1996}
\bibinfo{author}{De~Smet, C.}, \bibinfo{author}{De~Backer, O.},
  \bibinfo{author}{Faraoni, I.}, \bibinfo{author}{Lurquin, C.},
  \bibinfo{author}{Brasseur, F.}, \bibinfo{author}{Boon, T.},
  \bibinfo{year}{1996}.
\newblock \bibinfo{title}{The activation of human gene \uppercase{MAGE}-1 in
  tumor cells is correlated with genome-wide demethylation}.
\newblock \bibinfo{journal}{Proc Nat Acad Sci} \bibinfo{volume}{93},
  \bibinfo{pages}{7149--7153}.
\bibitem[{Delitala et~al.(2013)Delitala, Dianzani, Lorenzi and
  Melensi}]{Delitala2013a}
\bibinfo{author}{Delitala, M.}, \bibinfo{author}{Dianzani, U.},
  \bibinfo{author}{Lorenzi, T.}, \bibinfo{author}{Melensi, M.},
  \bibinfo{year}{2013}.
\newblock \bibinfo{title}{A mathematical model for immune and autoimmune
  response mediated by \uppercase{T}-cells}.
\newblock \bibinfo{journal}{Comput Math Appl} \bibinfo{volume}{66},
  \bibinfo{pages}{1010--1023}.
\bibitem[{Delitala and Lorenzi(2013)}]{Delitala2013}
\bibinfo{author}{Delitala, M.}, \bibinfo{author}{Lorenzi, T.},
  \bibinfo{year}{2013}.
\newblock \bibinfo{title}{Recognition and learning in a mathematical model for
  immune response against cancer.}
\newblock \bibinfo{journal}{Discrete Contin Dyn Syst Ser B}
  \bibinfo{volume}{18}, \bibinfo{pages}{891--914}.
\bibitem[{Deroulers et~al.(2009)Deroulers, Aubert, Badoual and
  Grammaticos}]{Deroulers2009}
\bibinfo{author}{Deroulers, C.}, \bibinfo{author}{Aubert, M.},
  \bibinfo{author}{Badoual, M.}, \bibinfo{author}{Grammaticos, B.},
  \bibinfo{year}{2009}.
\newblock \bibinfo{title}{Modeling tumor cell migration: \uppercase{F}rom
  microscopic to macroscopic models}.
\newblock \bibinfo{journal}{Phys Rev E} \bibinfo{volume}{79},
  \bibinfo{pages}{031917}.
\bibitem[{d'Onofrio and Ciancio(2011)}]{dOnofrio2011}
\bibinfo{author}{d'Onofrio, A.}, \bibinfo{author}{Ciancio, A.},
  \bibinfo{year}{2011}.
\newblock \bibinfo{title}{Simple biophysical model of tumor evasion from immune
  system control}.
\newblock \bibinfo{journal}{Phys Rev E} \bibinfo{volume}{84},
  \bibinfo{pages}{031910}.
\bibitem[{Ehrlich(2002)}]{Ehrlich2002}
\bibinfo{author}{Ehrlich, M.}, \bibinfo{year}{2002}.
\newblock \bibinfo{title}{\uppercase{DNA} methylation in cancer:
  \uppercase{T}oo much, but also too little}.
\newblock \bibinfo{journal}{Oncogene} \bibinfo{volume}{21},
  \bibinfo{pages}{5400}.
\bibitem[{Engelhardt et~al.(2012)Engelhardt, Boldajipour, Beemiller,
  Pandurangi, Sorensen, Werb, Egeblad and Krummel}]{Engelhardt2012}
\bibinfo{author}{Engelhardt, J.J.}, \bibinfo{author}{Boldajipour, B.},
  \bibinfo{author}{Beemiller, P.}, \bibinfo{author}{Pandurangi, P.},
  \bibinfo{author}{Sorensen, C.}, \bibinfo{author}{Werb, Z.},
  \bibinfo{author}{Egeblad, M.}, \bibinfo{author}{Krummel, M.F.},
  \bibinfo{year}{2012}.
\newblock \bibinfo{title}{Marginating dendritic cells of the tumor
  microenvironment cross-present tumor antigens and stably engage
  tumor-specific \uppercase{T} cells}.
\newblock \bibinfo{journal}{Cancer Cell} \bibinfo{volume}{21},
  \bibinfo{pages}{402--417}.
\bibitem[{Fehres et~al.(2014)Fehres, Unger, Garcia-Vallejo and van
  Kooyk}]{fehres2014}
\bibinfo{author}{Fehres, C.M.}, \bibinfo{author}{Unger, W.W.J.},
  \bibinfo{author}{Garcia-Vallejo, J.J.}, \bibinfo{author}{van Kooyk, Y.},
  \bibinfo{year}{2014}.
\newblock \bibinfo{title}{Understanding the biology of antigen
  cross-presentation for the design of vaccines against cancer}.
\newblock \bibinfo{journal}{Front Immunol} \bibinfo{volume}{5},
  \bibinfo{pages}{149}.
\bibitem[{Feinberg(2004)}]{Feinberg2004a}
\bibinfo{author}{Feinberg, A.P.}, \bibinfo{year}{2004}.
\newblock \bibinfo{title}{The epigenetics of cancer etiology}, in:
  \bibinfo{booktitle}{Seminars in Cancer Biology},
  \bibinfo{organization}{Elsevier}. pp. \bibinfo{pages}{427--432}.
\bibitem[{Feinberg and Tycko(2004)}]{Feinberg2004b}
\bibinfo{author}{Feinberg, A.P.}, \bibinfo{author}{Tycko, B.},
  \bibinfo{year}{2004}.
\newblock \bibinfo{title}{The history of cancer epigenetics}.
\newblock \bibinfo{journal}{Nat Rev Cancer} \bibinfo{volume}{4},
  \bibinfo{pages}{143}.
\bibitem[{Gerdemann et~al.(2011)Gerdemann, Katari, Christin, Cruz, Tripic,
  Rousseau, Gottschalk, Savoldo, Vera, Heslop et~al.}]{Gerdemann2011}
\bibinfo{author}{Gerdemann, U.}, \bibinfo{author}{Katari, U.},
  \bibinfo{author}{Christin, A.S.}, \bibinfo{author}{Cruz, C.R.},
  \bibinfo{author}{Tripic, T.}, \bibinfo{author}{Rousseau, A.},
  \bibinfo{author}{Gottschalk, S.M.}, \bibinfo{author}{Savoldo, B.},
  \bibinfo{author}{Vera, J.F.}, \bibinfo{author}{Heslop, H.E.}, et~al.,
  \bibinfo{year}{2011}.
\newblock \bibinfo{title}{Cytotoxic \uppercase{T} lymphocytes simultaneously
  targeting multiple tumor-associated antigens to treat \uppercase{EBV}
  negative lymphoma}.
\newblock \bibinfo{journal}{Mol Theory} \bibinfo{volume}{19},
  \bibinfo{pages}{2258--2268}.
\bibitem[{Gomis and Gawrzak(2017)}]{Gomis2017}
\bibinfo{author}{Gomis, R.R.}, \bibinfo{author}{Gawrzak, S.},
  \bibinfo{year}{2017}.
\newblock \bibinfo{title}{Tumor cell dormancy}.
\newblock \bibinfo{journal}{Mol Oncol} \bibinfo{volume}{11},
  \bibinfo{pages}{62--78}.
\bibitem[{Graff-Dubois et~al.(2002)Graff-Dubois, Faure, Gross, Alves, Scardino,
  Chouaib, Lemonnier and Kosmatopoulos}]{Graff2002}
\bibinfo{author}{Graff-Dubois, S.}, \bibinfo{author}{Faure, O.},
  \bibinfo{author}{Gross, D.A.}, \bibinfo{author}{Alves, P.},
  \bibinfo{author}{Scardino, A.}, \bibinfo{author}{Chouaib, S.},
  \bibinfo{author}{Lemonnier, F.A.}, \bibinfo{author}{Kosmatopoulos, K.},
  \bibinfo{year}{2002}.
\newblock \bibinfo{title}{Generation of \uppercase{CTL} recognizing an
  \uppercase{HLA-A* 0201}-restricted epitope shared by
  \uppercase{MAGE-A1,-A2,-A3,-A4,-A6,-A10}, and-\uppercase{A12} tumor antigens:
  Implication in a broad-spectrum tumor immunotherapy}.
\newblock \bibinfo{journal}{J Immunol} \bibinfo{volume}{169},
  \bibinfo{pages}{575--580}.
\bibitem[{Hanahan and Weinberg(2011)}]{Hanahan2011}
\bibinfo{author}{Hanahan, D.}, \bibinfo{author}{Weinberg, R.A.},
  \bibinfo{year}{2011}.
\newblock \bibinfo{title}{Hallmarks of cancer: The next generation}.
\newblock \bibinfo{journal}{Cell} \bibinfo{volume}{144},
  \bibinfo{pages}{646--674}.
\bibitem[{Harris et~al.(2012)Harris, Banigan, Christian, Konradt, Wojno,
  Norose, Wilson, John, Weninger, Luster et~al.}]{Harris2012}
\bibinfo{author}{Harris, T.H.}, \bibinfo{author}{Banigan, E.J.},
  \bibinfo{author}{Christian, D.A.}, \bibinfo{author}{Konradt, C.},
  \bibinfo{author}{Wojno, E.D.T.}, \bibinfo{author}{Norose, K.},
  \bibinfo{author}{Wilson, E.H.}, \bibinfo{author}{John, B.},
  \bibinfo{author}{Weninger, W.}, \bibinfo{author}{Luster, A.D.}, et~al.,
  \bibinfo{year}{2012}.
\newblock \bibinfo{title}{Generalized \uppercase{L}\'evy walks and the role of
  chemokines in migration of effector \uppercase{CD8+ T} cells}.
\newblock \bibinfo{journal}{Nature} \bibinfo{volume}{486},
  \bibinfo{pages}{545--548}.
\bibitem[{Hartmann et~al.(2016)Hartmann, Brisam, Rauthe, Driemel, Brands,
  Rosenwald, K{\"u}bler and M{\"u}ller-Richter}]{Hartmann2016}
\bibinfo{author}{Hartmann, S.}, \bibinfo{author}{Brisam, M.},
  \bibinfo{author}{Rauthe, S.}, \bibinfo{author}{Driemel, O.},
  \bibinfo{author}{Brands, R.C.}, \bibinfo{author}{Rosenwald, A.},
  \bibinfo{author}{K{\"u}bler, A.C.}, \bibinfo{author}{M{\"u}ller-Richter,
  U.D.}, \bibinfo{year}{2016}.
\newblock \bibinfo{title}{Contrary melanoma-associated antigen-\uppercase{A}
  expression at the tumor front and center: A comparative analysis of stage
  \uppercase{I} and \uppercase{IV} head and neck squamous cell carcinoma}.
\newblock \bibinfo{journal}{Oncol Lett} \bibinfo{volume}{12},
  \bibinfo{pages}{2942--2947}.
\bibitem[{Hartmann et~al.(2014)Hartmann, Kriegebaum, K{\"u}chler, Brands, Linz,
  K{\"u}bler and M{\"u}ller-Richter}]{hartmann2014}
\bibinfo{author}{Hartmann, S.}, \bibinfo{author}{Kriegebaum, U.},
  \bibinfo{author}{K{\"u}chler, N.}, \bibinfo{author}{Brands, R.C.},
  \bibinfo{author}{Linz, C.}, \bibinfo{author}{K{\"u}bler, A.C.},
  \bibinfo{author}{M{\"u}ller-Richter, U.D.A.}, \bibinfo{year}{2014}.
\newblock \bibinfo{title}{Correlation of \uppercase{MAGE-A} tumor antigens and
  the efficacy of various chemotherapeutic agents in head and neck carcinoma
  cells}.
\newblock \bibinfo{journal}{Clin Oral Investig} \bibinfo{volume}{18},
  \bibinfo{pages}{189--197}.
\bibitem[{Hartmann et~al.(2013)Hartmann, Kriegebaum, K{\"u}chler, Lessner,
  Brands, Linz, Schneider, K{\"u}bler and M{\"u}ller-Richter}]{hartmann2013}
\bibinfo{author}{Hartmann, S.}, \bibinfo{author}{Kriegebaum, U.},
  \bibinfo{author}{K{\"u}chler, N.}, \bibinfo{author}{Lessner, G.},
  \bibinfo{author}{Brands, R.C.}, \bibinfo{author}{Linz, C.},
  \bibinfo{author}{Schneider, T.}, \bibinfo{author}{K{\"u}bler, A.C.},
  \bibinfo{author}{M{\"u}ller-Richter, U.D.A.}, \bibinfo{year}{2013}.
\newblock \bibinfo{title}{Efficacy of cetuximab and panitumumab in oral
  squamous cell carcinoma cell lines: \uppercase{P}rognostic value of
  \uppercase{MAGE-A} subgroups for treatment success}.
\newblock \bibinfo{journal}{J Cranio Maxill Surg} \bibinfo{volume}{41},
  \bibinfo{pages}{623--629}.
\bibitem[{Johnston et~al.(2007)Johnston, Edwards, Bodmer, Maini and
  Chapman}]{Johnston2007}
\bibinfo{author}{Johnston, M.D.}, \bibinfo{author}{Edwards, C.M.},
  \bibinfo{author}{Bodmer, W.F.}, \bibinfo{author}{Maini, P.K.},
  \bibinfo{author}{Chapman, S.J.}, \bibinfo{year}{2007}.
\newblock \bibinfo{title}{Mathematical modeling of cell population dynamics in
  the colonic crypt and in colorectal cancer}.
\newblock \bibinfo{journal}{Proc Nat Acad Sci} \bibinfo{volume}{104},
  \bibinfo{pages}{4008--4013}.
\bibitem[{June et~al.(2018)June, O'Connor, Kawalekar, Ghassemi and
  Milone}]{june2018}
\bibinfo{author}{June, C.H.}, \bibinfo{author}{O'Connor, R.S.},
  \bibinfo{author}{Kawalekar, O.U.}, \bibinfo{author}{Ghassemi, S.},
  \bibinfo{author}{Milone, M.C.}, \bibinfo{year}{2018}.
\newblock \bibinfo{title}{\uppercase{CAR T} cell immunotherapy for human
  cancer}.
\newblock \bibinfo{journal}{Science} \bibinfo{volume}{359},
  \bibinfo{pages}{1361--1365}.
\bibitem[{Kenney and Keeping(1962)}]{stats}
\bibinfo{author}{Kenney, J.F.}, \bibinfo{author}{Keeping, E.S.},
  \bibinfo{year}{1962}.
\newblock \bibinfo{title}{Mathematics of Statistics Pt. 1}.
\newblock \bibinfo{publisher}{Princeton, NJ: Van Nostrand}.
\bibitem[{Kolev et~al.(2013)Kolev, Nawrocki and Zubik-Kowal}]{Kolev2013}
\bibinfo{author}{Kolev, M.}, \bibinfo{author}{Nawrocki, S.},
  \bibinfo{author}{Zubik-Kowal, B.}, \bibinfo{year}{2013}.
\newblock \bibinfo{title}{Numerical simulations for tumor and cellular immune
  system interactions in lung cancer treatment}.
\newblock \bibinfo{journal}{Commun Nonlinear Sci Numer Simul}
  \bibinfo{volume}{18}, \bibinfo{pages}{1473--1480}.
\bibitem[{K{\"o}se et~al.(2017)K{\"o}se, Moore, Ofodile, Radunskaya, Swanson
  and Zollinger}]{Kose2017}
\bibinfo{author}{K{\"o}se, E.}, \bibinfo{author}{Moore, S.},
  \bibinfo{author}{Ofodile, C.}, \bibinfo{author}{Radunskaya, A.},
  \bibinfo{author}{Swanson, E.R.}, \bibinfo{author}{Zollinger, E.},
  \bibinfo{year}{2017}.
\newblock \bibinfo{title}{Immuno-kinetics of immunotherapy: \uppercase{D}osing
  with \uppercase{DC}s}.
\newblock \bibinfo{journal}{Lett Biomath} \bibinfo{volume}{4},
  \bibinfo{pages}{39--58}.
\bibitem[{Kuznetsov et~al.(1994)Kuznetsov, Makalkin, Taylor and
  Perelson}]{Kuznetsov1994}
\bibinfo{author}{Kuznetsov, V.A.}, \bibinfo{author}{Makalkin, I.A.},
  \bibinfo{author}{Taylor, M.A.}, \bibinfo{author}{Perelson, A.S.},
  \bibinfo{year}{1994}.
\newblock \bibinfo{title}{Nonlinear dynamics of immunogenic tumors: Parameter
  estimation and global bifurcation analysis}.
\newblock \bibinfo{journal}{Bull Math Biol} \bibinfo{volume}{56},
  \bibinfo{pages}{295--321}.
\bibitem[{Linette et~al.(2013)Linette, Stadtmauer, Maus, Rapoport, Levine,
  Emery, Litzky, Bagg, Carreno, Cimino et~al.}]{linette2013}
\bibinfo{author}{Linette, G.P.}, \bibinfo{author}{Stadtmauer, E.A.},
  \bibinfo{author}{Maus, M.V.}, \bibinfo{author}{Rapoport, A.P.},
  \bibinfo{author}{Levine, B.L.}, \bibinfo{author}{Emery, L.},
  \bibinfo{author}{Litzky, L.}, \bibinfo{author}{Bagg, A.},
  \bibinfo{author}{Carreno, B.M.}, \bibinfo{author}{Cimino, P.J.}, et~al.,
  \bibinfo{year}{2013}.
\newblock \bibinfo{title}{Cardiovascular toxicity and titin cross-reactivity of
  affinity-enhanced \uppercase{T} cells in myeloma and melanoma}.
\newblock \bibinfo{journal}{Blood} \bibinfo{volume}{122},
  \bibinfo{pages}{863--871}.
\bibitem[{Ljunggren et~al.(1990)Ljunggren, Stam, {\"O}hl{\'e}n, Neefjes,
  H{\"o}glund, Heemels, Bastin, Schumacher, Townsend, K{\"a}rre
  et~al.}]{ljunggren1990}
\bibinfo{author}{Ljunggren, H.}, \bibinfo{author}{Stam, N.J.},
  \bibinfo{author}{{\"O}hl{\'e}n, C.}, \bibinfo{author}{Neefjes, J.J.},
  \bibinfo{author}{H{\"o}glund, P.}, \bibinfo{author}{Heemels, M.},
  \bibinfo{author}{Bastin, J.}, \bibinfo{author}{Schumacher, T.N.M.},
  \bibinfo{author}{Townsend, A.}, \bibinfo{author}{K{\"a}rre, K.}, et~al.,
  \bibinfo{year}{1990}.
\newblock \bibinfo{title}{Empty \uppercase{MHC} class \uppercase{I} molecules
  come out in the cold}.
\newblock \bibinfo{journal}{Nature} \bibinfo{volume}{346},
  \bibinfo{pages}{476}.
\bibitem[{Lorenzi et~al.(2016)Lorenzi, Chisholm and Clairambault}]{Lorenzi2016}
\bibinfo{author}{Lorenzi, T.}, \bibinfo{author}{Chisholm, R.H.},
  \bibinfo{author}{Clairambault, J.}, \bibinfo{year}{2016}.
\newblock \bibinfo{title}{Tracking the evolution of cancer cell populations
  through the mathematical lens of phenotype-structured equations}.
\newblock \bibinfo{journal}{Biol Direct} \bibinfo{volume}{11},
  \bibinfo{pages}{43}.
\bibitem[{Lorenzi et~al.(2015)Lorenzi, Chisholm, Melensi, Lorz and
  Delitala}]{Lorenzi2015}
\bibinfo{author}{Lorenzi, T.}, \bibinfo{author}{Chisholm, R.H.},
  \bibinfo{author}{Melensi, M.}, \bibinfo{author}{Lorz, A.},
  \bibinfo{author}{Delitala, M.}, \bibinfo{year}{2015}.
\newblock \bibinfo{title}{Mathematical model reveals how regulating the three
  phases of \uppercase{T}-cell response could counteract immune evasion}.
\newblock \bibinfo{journal}{Immunology} \bibinfo{volume}{146},
  \bibinfo{pages}{271--280}.
\bibitem[{{\L}uksza et~al.(2017){\L}uksza, Riaz, Makarov, Balachandran,
  Hellmann, Solovyov, Rizvi, Merghoub, Levine, Chan et~al.}]{Luksza2017}
\bibinfo{author}{{\L}uksza, M.}, \bibinfo{author}{Riaz, N.},
  \bibinfo{author}{Makarov, V.}, \bibinfo{author}{Balachandran, V.P.},
  \bibinfo{author}{Hellmann, M.D.}, \bibinfo{author}{Solovyov, A.},
  \bibinfo{author}{Rizvi, N.A.}, \bibinfo{author}{Merghoub, T.},
  \bibinfo{author}{Levine, A.J.}, \bibinfo{author}{Chan, T.A.}, et~al.,
  \bibinfo{year}{2017}.
\newblock \bibinfo{title}{A neoantigen fitness model predicts tumour response
  to checkpoint blockade immunotherapy}.
\newblock \bibinfo{journal}{Nature} \bibinfo{volume}{551},
  \bibinfo{pages}{517--520}.
\bibitem[{Macfarlane et~al.(2018)Macfarlane, Lorenzi and
  Chaplain}]{Macfarlane2018}
\bibinfo{author}{Macfarlane, F.R.}, \bibinfo{author}{Lorenzi, T.},
  \bibinfo{author}{Chaplain, M.A.J.}, \bibinfo{year}{2018}.
\newblock \bibinfo{title}{Modelling the immune response to cancer: An
  individual-based approach accounting for the difference in movement between
  inactive and activated \uppercase{T} cells}.
\newblock \bibinfo{journal}{Bull Math Biol} \bibinfo{volume}{80},
  \bibinfo{pages}{1539--1564}.
\bibitem[{Mallet and de~Pillis(2006)}]{Mallet2006}
\bibinfo{author}{Mallet, D.G.}, \bibinfo{author}{de~Pillis, L.G.},
  \bibinfo{year}{2006}.
\newblock \bibinfo{title}{A cellular automata model of tumor-immune system
  interactions}.
\newblock \bibinfo{journal}{J Theor Biol} \bibinfo{volume}{239},
  \bibinfo{pages}{334--350}.
\bibitem[{Manem et~al.(2014)Manem, Kohandel, Komarova and
  Sivaloganathan}]{Manem2014}
\bibinfo{author}{Manem, V.S.}, \bibinfo{author}{Kohandel, M.},
  \bibinfo{author}{Komarova, N.}, \bibinfo{author}{Sivaloganathan, S.},
  \bibinfo{year}{2014}.
\newblock \bibinfo{title}{Spatial invasion dynamics on random and unstructured
  meshes: \uppercase{I}mplications for heterogeneous tumor populations}.
\newblock \bibinfo{journal}{J Theor Biol} \bibinfo{volume}{349},
  \bibinfo{pages}{66--73}.
\bibitem[{Manjili(2018)}]{Manjili2018}
\bibinfo{author}{Manjili, M.H.}, \bibinfo{year}{2018}.
\newblock \bibinfo{title}{A theoretical basis for the efficacy of cancer
  immunotherapy and immunogenic tumor dormancy: The adaptation model of
  immunity}.
\newblock \bibinfo{journal}{Adv Cancer Res} \bibinfo{volume}{137},
  \bibinfo{pages}{17--36}.
\bibitem[{Marcar et~al.(2010)Marcar, MacLaine, Hupp and Meek}]{marcar2010}
\bibinfo{author}{Marcar, L.}, \bibinfo{author}{MacLaine, N.J.},
  \bibinfo{author}{Hupp, T.R.}, \bibinfo{author}{Meek, D.W.},
  \bibinfo{year}{2010}.
\newblock \bibinfo{title}{\uppercase{MAGE-A} cancer/testis antigens inhibit p53
  function by blocking its interaction with chromatin}.
\newblock \bibinfo{journal}{Cancer Res} \bibinfo{volume}{70},
  \bibinfo{pages}{10362--10370}.
\bibitem[{Matzavinos and Chaplain(2004)}]{Matzavinos2004b}
\bibinfo{author}{Matzavinos, A.}, \bibinfo{author}{Chaplain, M.A.J.},
  \bibinfo{year}{2004}.
\newblock \bibinfo{title}{Travelling-wave analysis of a model of the immune
  response to cancer}.
\newblock \bibinfo{journal}{C R Biol} \bibinfo{volume}{327},
  \bibinfo{pages}{995--1008}.
\bibitem[{Matzavinos et~al.(2004)Matzavinos, Chaplain and
  Kuznetsov}]{Matzavinos2004}
\bibinfo{author}{Matzavinos, A.}, \bibinfo{author}{Chaplain, M.A.J.},
  \bibinfo{author}{Kuznetsov, V.A.}, \bibinfo{year}{2004}.
\newblock \bibinfo{title}{Mathematical modelling of the spatio-temporal
  response of cytotoxic \uppercase{T}-lymphocytes to a solid tumour}.
\newblock \bibinfo{journal}{Math Med Biol} \bibinfo{volume}{21},
  \bibinfo{pages}{1--34}.
\bibitem[{Mellman et~al.(2011)Mellman, Coukos and Dranoff}]{mellman2011}
\bibinfo{author}{Mellman, I.}, \bibinfo{author}{Coukos, G.},
  \bibinfo{author}{Dranoff, G.}, \bibinfo{year}{2011}.
\newblock \bibinfo{title}{Cancer immunotherapy comes of age}.
\newblock \bibinfo{journal}{Nature} \bibinfo{volume}{480},
  \bibinfo{pages}{480}.
\bibitem[{Messerschmidt et~al.(2016)Messerschmidt, Prendergast and
  Messerschmidt}]{Messerschmidt2016}
\bibinfo{author}{Messerschmidt, J.L.}, \bibinfo{author}{Prendergast, G.C.},
  \bibinfo{author}{Messerschmidt, G.L.}, \bibinfo{year}{2016}.
\newblock \bibinfo{title}{How cancers escape immune destruction and mechanisms
  of action for the new significantly active immune therapies: Helping
  non-immunologists decipher recent advances.}
\newblock \bibinfo{journal}{Oncologist} \bibinfo{volume}{21},
  \bibinfo{pages}{233--243}.
\bibitem[{Monte et~al.(2006)Monte, Simonatto, Peche, Bublik, Gobessi, Pierotti,
  Rodolfo and Schneider}]{monte2006}
\bibinfo{author}{Monte, M.}, \bibinfo{author}{Simonatto, M.},
  \bibinfo{author}{Peche, L.Y.}, \bibinfo{author}{Bublik, D.R.},
  \bibinfo{author}{Gobessi, S.}, \bibinfo{author}{Pierotti, M.A.},
  \bibinfo{author}{Rodolfo, M.}, \bibinfo{author}{Schneider, C.},
  \bibinfo{year}{2006}.
\newblock \bibinfo{title}{\uppercase{MAGE-A} tumor antigens target p53
  transactivation function through histone deacetylase recruitment and confer
  resistance to chemotherapeutic agents}.
\newblock \bibinfo{journal}{Proc Nat Acad Sci} \bibinfo{volume}{103},
  \bibinfo{pages}{11160--11165}.
\bibitem[{M{\"u}ller-Richter et~al.(2009)M{\"u}ller-Richter, Dowejko, Reuther,
  Kleinheinz, Reichert and Driemel}]{Muller2009}
\bibinfo{author}{M{\"u}ller-Richter, U.D.A.}, \bibinfo{author}{Dowejko, A.},
  \bibinfo{author}{Reuther, T.}, \bibinfo{author}{Kleinheinz, J.},
  \bibinfo{author}{Reichert, T.E.}, \bibinfo{author}{Driemel, O.},
  \bibinfo{year}{2009}.
\newblock \bibinfo{title}{Analysis of expression profiles of \uppercase{MAGE-A}
  antigens in oral squamous cell carcinoma cell lines}.
\newblock \bibinfo{journal}{Head Face Med} \bibinfo{volume}{5},
  \bibinfo{pages}{10}.
\bibitem[{Oey and Whitelaw(2014)}]{Oey2014}
\bibinfo{author}{Oey, H.}, \bibinfo{author}{Whitelaw, E.},
  \bibinfo{year}{2014}.
\newblock \bibinfo{title}{On the meaning of the word ``epimutation"}.
\newblock \bibinfo{journal}{Trends Genet} \bibinfo{volume}{30},
  \bibinfo{pages}{519--520}.
\bibitem[{Ott et~al.(2017)Ott, Hu, Keskin, Shuklka, Sun et~al.}]{Ott2017}
\bibinfo{author}{Ott, P.A.}, \bibinfo{author}{Hu, Z.}, \bibinfo{author}{Keskin,
  D.B.}, \bibinfo{author}{Shuklka, S.A.}, \bibinfo{author}{Sun, J.}, et~al.,
  \bibinfo{year}{2017}.
\newblock \bibinfo{title}{An immunogenic personal neoantigen vaccine for
  patients with melanoma}.
\newblock \bibinfo{journal}{Nature} \bibinfo{volume}{547},
  \bibinfo{pages}{217--221}.
\bibitem[{Painter and Hillen(2015)}]{Painter2015}
\bibinfo{author}{Painter, K.J.}, \bibinfo{author}{Hillen, T.},
  \bibinfo{year}{2015}.
\newblock \bibinfo{title}{Navigating the flow: \uppercase{I}ndividual and
  continuum models for homing in flowing environments}.
\newblock \bibinfo{journal}{J R Soc Interface} \bibinfo{volume}{12},
  \bibinfo{pages}{20150647}.
\bibitem[{Peltom{\"a}ki(2012)}]{Peltomaki2012}
\bibinfo{author}{Peltom{\"a}ki, P.}, \bibinfo{year}{2012}.
\newblock \bibinfo{title}{Mutations and epimutations in the origin of cancer}.
\newblock \bibinfo{journal}{Exp Cell Res} \bibinfo{volume}{318},
  \bibinfo{pages}{299--310}.
\bibitem[{Penington et~al.(2011)Penington, Hughes and Landman}]{Penington2011}
\bibinfo{author}{Penington, C.J.}, \bibinfo{author}{Hughes, B.D.},
  \bibinfo{author}{Landman, K.A.}, \bibinfo{year}{2011}.
\newblock \bibinfo{title}{Building macroscale models from microscale
  probabilistic models: a general probabilistic approach for nonlinear
  diffusion and multispecies phenomena}.
\newblock \bibinfo{journal}{Phys Rev E} \bibinfo{volume}{84},
  \bibinfo{pages}{041120}.
\bibitem[{de~Pillis et~al.(2009)de~Pillis, Renee~Fister, Gu, Collins, Daub,
  Gross, Moore and Preskill}]{dePillis2009}
\bibinfo{author}{de~Pillis, L.}, \bibinfo{author}{Renee~Fister, K.},
  \bibinfo{author}{Gu, W.}, \bibinfo{author}{Collins, C.},
  \bibinfo{author}{Daub, M.}, \bibinfo{author}{Gross, D.},
  \bibinfo{author}{Moore, J.}, \bibinfo{author}{Preskill, B.},
  \bibinfo{year}{2009}.
\newblock \bibinfo{title}{Mathematical model creation for cancer
  chemo-immunotherapy}.
\newblock \bibinfo{journal}{Comput Math Meth Med} \bibinfo{volume}{10},
  \bibinfo{pages}{165--184}.
\bibitem[{de~Pillis et~al.(2006)de~Pillis, Mallet and
  Radunskaya}]{dePillis2006}
\bibinfo{author}{de~Pillis, L.G.}, \bibinfo{author}{Mallet, D.G.},
  \bibinfo{author}{Radunskaya, A.E.}, \bibinfo{year}{2006}.
\newblock \bibinfo{title}{Spatial tumor-immune modeling}.
\newblock \bibinfo{journal}{Comput Math Meth Med} \bibinfo{volume}{7},
  \bibinfo{pages}{159--176}.
\bibitem[{Raman et~al.(2016)Raman, Rizkallah, Simmons, Donnellan, Dukes, Bossi,
  Le~Provost, Todorov, Baston, Hickman et~al.}]{raman2016}
\bibinfo{author}{Raman, M.C.C.}, \bibinfo{author}{Rizkallah, P.J.},
  \bibinfo{author}{Simmons, R.}, \bibinfo{author}{Donnellan, Z.},
  \bibinfo{author}{Dukes, J.}, \bibinfo{author}{Bossi, G.},
  \bibinfo{author}{Le~Provost, G.S.}, \bibinfo{author}{Todorov, P.},
  \bibinfo{author}{Baston, E.}, \bibinfo{author}{Hickman, E.}, et~al.,
  \bibinfo{year}{2016}.
\newblock \bibinfo{title}{Direct molecular mimicry enables off-target
  cardiovascular toxicity by an enhanced affinity \uppercase{TCR} designed for
  cancer immunotherapy}.
\newblock \bibinfo{journal}{Sci Rep} \bibinfo{volume}{6},
  \bibinfo{pages}{18851}.
\bibitem[{Ribas and Wolchok(2018)}]{ribas2018}
\bibinfo{author}{Ribas, A.}, \bibinfo{author}{Wolchok, J.D.},
  \bibinfo{year}{2018}.
\newblock \bibinfo{title}{Cancer immunotherapy using checkpoint blockade}.
\newblock \bibinfo{journal}{Science} \bibinfo{volume}{349},
  \bibinfo{pages}{1350--1355}.
\bibitem[{Roch et~al.(2010)Roch, Kutup, Vashist, Yekebas, Kalinin and
  Izbicki}]{roch2010}
\bibinfo{author}{Roch, N.}, \bibinfo{author}{Kutup, A.},
  \bibinfo{author}{Vashist, Y.}, \bibinfo{author}{Yekebas, E.},
  \bibinfo{author}{Kalinin, V.}, \bibinfo{author}{Izbicki, J.R.},
  \bibinfo{year}{2010}.
\newblock \bibinfo{title}{Coexpression of \uppercase{MAGE-A} peptides and
  \uppercase{HLA} class \uppercase{I} molecules in hepatocellular carcinoma}.
\newblock \bibinfo{journal}{Anticancer Res} \bibinfo{volume}{30},
  \bibinfo{pages}{1617--1623}.
\bibitem[{Sahin et~al.(2017)Sahin, Derhovanessian, Miller, Kloke, Simon
  et~al.}]{Sahin2017}
\bibinfo{author}{Sahin, U.}, \bibinfo{author}{Derhovanessian, E.},
  \bibinfo{author}{Miller, M.}, \bibinfo{author}{Kloke, B.},
  \bibinfo{author}{Simon, P.}, et~al., \bibinfo{year}{2017}.
\newblock \bibinfo{title}{Personalized \uppercase{RNA} mutanome vaccines
  mobilize poly-specific therapeutic immunity against cancer}.
\newblock \bibinfo{journal}{Nature} \bibinfo{volume}{547},
  \bibinfo{pages}{222--226}.
\bibitem[{Schmid et~al.(2010)Schmid, Irving, Posevitz, Hebeisen,
  Posevitz-Fejfar, Sarria, Gomez-Eerland, Thome, Schumacher, Romero
  et~al.}]{schmid2010}
\bibinfo{author}{Schmid, D.A.}, \bibinfo{author}{Irving, M.B.},
  \bibinfo{author}{Posevitz, V.}, \bibinfo{author}{Hebeisen, M.},
  \bibinfo{author}{Posevitz-Fejfar, A.}, \bibinfo{author}{Sarria, J.C.F.},
  \bibinfo{author}{Gomez-Eerland, R.}, \bibinfo{author}{Thome, M.},
  \bibinfo{author}{Schumacher, T.N.M.}, \bibinfo{author}{Romero, P.}, et~al.,
  \bibinfo{year}{2010}.
\newblock \bibinfo{title}{Evidence for a \uppercase{TCR} affinity threshold
  delimiting maximal \uppercase{CD8 T} cell function}.
\newblock \bibinfo{journal}{J Immunol} \bibinfo{volume}{184},
  \bibinfo{pages}{4936--4946}.
\bibitem[{Schueler-Furman et~al.(1998)Schueler-Furman, Elber and
  Margalit}]{schueler1998}
\bibinfo{author}{Schueler-Furman, O.}, \bibinfo{author}{Elber, R.},
  \bibinfo{author}{Margalit, H.}, \bibinfo{year}{1998}.
\newblock \bibinfo{title}{Knowledge-based structure prediction of
  \uppercase{MHC} class \uppercase{I} bound peptides: \uppercase{A} study of 23
  complexes}.
\newblock \bibinfo{journal}{Fold Des} \bibinfo{volume}{3},
  \bibinfo{pages}{549--564}.
\bibitem[{Schumacher and Hacohen(2016)}]{Schumacher2016}
\bibinfo{author}{Schumacher, T.N.}, \bibinfo{author}{Hacohen, N.},
  \bibinfo{year}{2016}.
\newblock \bibinfo{title}{Neoantigens encoded in the cancer genome}.
\newblock \bibinfo{journal}{Curr Opinion Immunol} \bibinfo{volume}{41},
  \bibinfo{pages}{98--103}.
\bibitem[{Sharma and Allison(2015)}]{Sharma2015}
\bibinfo{author}{Sharma, P.}, \bibinfo{author}{Allison, J.P.},
  \bibinfo{year}{2015}.
\newblock \bibinfo{title}{The future of immune checkpoint therapy}.
\newblock \bibinfo{journal}{Science} \bibinfo{volume}{348},
  \bibinfo{pages}{56--61}.
\bibitem[{Slansky and Jordan(2010)}]{slansky2010}
\bibinfo{author}{Slansky, J.E.}, \bibinfo{author}{Jordan, K.R.},
  \bibinfo{year}{2010}.
\newblock \bibinfo{title}{The \uppercase{G}oldilocks model for \uppercase{TCR}:
  \uppercase{T}oo much attraction might not be best for vaccine design}.
\newblock \bibinfo{journal}{PLoS Biol} \bibinfo{volume}{8},
  \bibinfo{pages}{e1000482}.
\bibitem[{Smith et~al.(2014)Smith, Wang, Baylon, Singh, Baker, Tajkhorshid and
  Kranz}]{smith2014}
\bibinfo{author}{Smith, S.N.}, \bibinfo{author}{Wang, Y.},
  \bibinfo{author}{Baylon, J.L.}, \bibinfo{author}{Singh, N.K.},
  \bibinfo{author}{Baker, B.M.}, \bibinfo{author}{Tajkhorshid, E.},
  \bibinfo{author}{Kranz, D.M.}, \bibinfo{year}{2014}.
\newblock \bibinfo{title}{Changing the peptide specificity of a human
  \uppercase{T} cell receptor by directed evolution}.
\newblock \bibinfo{journal}{Nature Comm} \bibinfo{volume}{5},
  \bibinfo{pages}{5223}.
\bibitem[{Stace et~al.(2019)Stace, Stiehl, Marciniak-Czochra and
  Lorenzi}]{stace2019}
\bibinfo{author}{Stace, R.E.}, \bibinfo{author}{Stiehl, T.},
  \bibinfo{author}{Marciniak-Czochra, A.}, \bibinfo{author}{Lorenzi, T.},
  \bibinfo{year}{2019}.
\newblock \bibinfo{title}{A phenotype-structured individual-based model for the
  evolution of cancer cell populations under chemotherapy}.
\newblock \bibinfo{journal}{Preprint} .
\bibitem[{Stevens(2000)}]{Stevens2000}
\bibinfo{author}{Stevens, A.}, \bibinfo{year}{2000}.
\newblock \bibinfo{title}{The derivation of chemotaxis equations as limit
  dynamics of moderately interacting stochastic many-particle systems}.
\newblock \bibinfo{journal}{SIAM J Appl Math} \bibinfo{volume}{61},
  \bibinfo{pages}{183--212}.
\bibitem[{Stone et~al.(2009)Stone, Chervin and Kranz}]{Stone2009}
\bibinfo{author}{Stone, J.D.}, \bibinfo{author}{Chervin, A.S.},
  \bibinfo{author}{Kranz, D.M.}, \bibinfo{year}{2009}.
\newblock \bibinfo{title}{T-cell receptor binding affinities and kinetics:
  \uppercase{I}mpact on \uppercase{T}-cell activity and specificity}.
\newblock \bibinfo{journal}{Immunology} \bibinfo{volume}{126},
  \bibinfo{pages}{165--176}.
\bibitem[{Tan et~al.(2015)Tan, Gerry, Brewer, Melchiori, Bridgeman, Bennett,
  Pumphrey, Jakobsen, Price, Ladell et~al.}]{Tan2015}
\bibinfo{author}{Tan, M.P.}, \bibinfo{author}{Gerry, A.B.},
  \bibinfo{author}{Brewer, J.E.}, \bibinfo{author}{Melchiori, L.},
  \bibinfo{author}{Bridgeman, J.S.}, \bibinfo{author}{Bennett, A.D.},
  \bibinfo{author}{Pumphrey, N.J.}, \bibinfo{author}{Jakobsen, B.K.},
  \bibinfo{author}{Price, D.A.}, \bibinfo{author}{Ladell, K.}, et~al.,
  \bibinfo{year}{2015}.
\newblock \bibinfo{title}{T cell receptor binding affinity governs the
  functional profile of cancer-specific \uppercase{CD8}+ \uppercase{T} cells}.
\newblock \bibinfo{journal}{Clin Exp Immunol} \bibinfo{volume}{180},
  \bibinfo{pages}{255--270}.
\bibitem[{Teng et~al.(2008)Teng, Swann, Koebel, Schreiber and Smyth}]{Teng2008}
\bibinfo{author}{Teng, M.W.L.}, \bibinfo{author}{Swann, J.B.},
  \bibinfo{author}{Koebel, C.M.}, \bibinfo{author}{Schreiber, R.D.},
  \bibinfo{author}{Smyth, M.J.}, \bibinfo{year}{2008}.
\newblock \bibinfo{title}{Immune-mediated dormancy: \uppercase{A}n equilibrium
  with cancer}.
\newblock \bibinfo{journal}{J Leukocyte Biol} \bibinfo{volume}{84},
  \bibinfo{pages}{988--993}.
\bibitem[{Tomasetti and Levy(2010)}]{Tomasetti2010}
\bibinfo{author}{Tomasetti, C.}, \bibinfo{author}{Levy, D.},
  \bibinfo{year}{2010}.
\newblock \bibinfo{title}{An elementary approach to modeling drug resistance in
  cancer}.
\newblock \bibinfo{journal}{Math Biosci Eng: MBE} \bibinfo{volume}{7},
  \bibinfo{pages}{905}.
\bibitem[{Tong et~al.(2004)Tong, Tan and Ranganathan}]{tong2004}
\bibinfo{author}{Tong, J.C.}, \bibinfo{author}{Tan, T.W.},
  \bibinfo{author}{Ranganathan, S.}, \bibinfo{year}{2004}.
\newblock \bibinfo{title}{Modeling the structure of bound peptide ligands to
  major histocompatibility complex}.
\newblock \bibinfo{journal}{Protein Sci} \bibinfo{volume}{13},
  \bibinfo{pages}{2523--2532}.
\bibitem[{Urosevic et~al.(2005)Urosevic, Braun, Willers, Burg and
  Dummer}]{Urosevic2005}
\bibinfo{author}{Urosevic, M.}, \bibinfo{author}{Braun, B.},
  \bibinfo{author}{Willers, J.}, \bibinfo{author}{Burg, G.},
  \bibinfo{author}{Dummer, R.}, \bibinfo{year}{2005}.
\newblock \bibinfo{title}{Expression of melanoma-associated antigens in
  melanoma cell cultures}.
\newblock \bibinfo{journal}{Exp Dermatol} \bibinfo{volume}{14},
  \bibinfo{pages}{491--497}.
\bibitem[{Van~Tongelen et~al.(2017)Van~Tongelen, Loriot and
  De~Smet}]{VanTongelen2017}
\bibinfo{author}{Van~Tongelen, A.}, \bibinfo{author}{Loriot, A.},
  \bibinfo{author}{De~Smet, C.}, \bibinfo{year}{2017}.
\newblock \bibinfo{title}{Oncogenic roles of \uppercase{DNA} hypomethylation
  through the activation of cancer-germline genes}.
\newblock \bibinfo{journal}{Cancer Lett} \bibinfo{volume}{396},
  \bibinfo{pages}{130--137}.
\bibitem[{Wang and Lin(2013)}]{Wang2013}
\bibinfo{author}{Wang, S.}, \bibinfo{author}{Lin, S.}, \bibinfo{year}{2013}.
\newblock \bibinfo{title}{Tumor dormancy: \uppercase{P}otential therapeutic
  target in tumor recurrence and metastasis prevention}.
\newblock \bibinfo{journal}{Exp Hematol Oncol} \bibinfo{volume}{2},
  \bibinfo{pages}{29}.
\bibitem[{Wischnewski et~al.(2006)Wischnewski, Pantel and
  Schwarzenbach}]{Wischnewski2006}
\bibinfo{author}{Wischnewski, F.}, \bibinfo{author}{Pantel, K.},
  \bibinfo{author}{Schwarzenbach, H.}, \bibinfo{year}{2006}.
\newblock \bibinfo{title}{Promoter demethylation and histone acetylation
  mediate gene expression of \uppercase{MAGE-A1,-A2,-A3}, and-\uppercase{A12}
  in human cancer cells}.
\newblock \bibinfo{journal}{Mol Cancer Res} \bibinfo{volume}{4},
  \bibinfo{pages}{339--349}.
\bibitem[{Wu et~al.(2018)Wu, Liao, Kirilin, Lin, Torga, Qu, Liu, Sturm, Pienta
  and Austin}]{Wu2018}
\bibinfo{author}{Wu, A.}, \bibinfo{author}{Liao, D.}, \bibinfo{author}{Kirilin,
  V.}, \bibinfo{author}{Lin, K.}, \bibinfo{author}{Torga, G.},
  \bibinfo{author}{Qu, J.}, \bibinfo{author}{Liu, L.}, \bibinfo{author}{Sturm,
  J.C.}, \bibinfo{author}{Pienta, K.}, \bibinfo{author}{Austin, R.},
  \bibinfo{year}{2018}.
\newblock \bibinfo{title}{Cancer dormancy and criticality from a game theory
  perspective}.
\newblock \bibinfo{journal}{Cancer Convergence} \bibinfo{volume}{2},
  \bibinfo{pages}{1}.
\bibitem[{Yang et~al.(2007)Yang, O'Herrin, Wu, Reagan-Shaw, Ma, Bhat,
  Gravekamp, Setaluri, Peters, Hoffmann et~al.}]{yang2007}
\bibinfo{author}{Yang, B.}, \bibinfo{author}{O'Herrin, S.M.},
  \bibinfo{author}{Wu, J.}, \bibinfo{author}{Reagan-Shaw, S.},
  \bibinfo{author}{Ma, Y.}, \bibinfo{author}{Bhat, K.M.R.},
  \bibinfo{author}{Gravekamp, C.}, \bibinfo{author}{Setaluri, V.},
  \bibinfo{author}{Peters, N.}, \bibinfo{author}{Hoffmann, F.M.}, et~al.,
  \bibinfo{year}{2007}.
\newblock \bibinfo{title}{\uppercase{MAGE-A, MAGE-B} and \uppercase{MAGE-C}
  proteins form complexes with \uppercase{KAP1} and suppress p53-dependent
  apoptosis in \uppercase{MAGE}-positive cell lines}.
\newblock \bibinfo{journal}{Cancer Res} \bibinfo{volume}{67},
  \bibinfo{pages}{9954--9962}.
\bibitem[{Yarchoan et~al.(2017)Yarchoan, Johnson, Lutz, Laheru and
  Jaffee}]{Yarchoan2017}
\bibinfo{author}{Yarchoan, M.}, \bibinfo{author}{Johnson, B.A.},
  \bibinfo{author}{Lutz, E.R.}, \bibinfo{author}{Laheru, D.A.},
  \bibinfo{author}{Jaffee, E.M.}, \bibinfo{year}{2017}.
\newblock \bibinfo{title}{Targeting neoantigens to augment antitumour
  immunity}.
\newblock \bibinfo{journal}{Nat Rev Cancer} \bibinfo{volume}{17},
  \bibinfo{pages}{209--222}.
\bibitem[{Yeh and Ramaswamy(2015)}]{Yeh2015}
\bibinfo{author}{Yeh, A.C.}, \bibinfo{author}{Ramaswamy, S.},
  \bibinfo{year}{2015}.
\newblock \bibinfo{title}{Mechanisms of cancer cell dormancy:
  \uppercase{A}nother hallmark of cancer?}
\newblock \bibinfo{journal}{Cancer Res} \bibinfo{volume}{75},
  \bibinfo{pages}{5014--5022}.
\bibitem[{Zajac et~al.(2017)Zajac, Schultz-Thater, Tornillo, Sadowski, Trella,
  Mengus, Iezzi and Spagnoli}]{Zajac2017}
\bibinfo{author}{Zajac, P.}, \bibinfo{author}{Schultz-Thater, E.},
  \bibinfo{author}{Tornillo, L.}, \bibinfo{author}{Sadowski, C.},
  \bibinfo{author}{Trella, E.}, \bibinfo{author}{Mengus, C.},
  \bibinfo{author}{Iezzi, G.}, \bibinfo{author}{Spagnoli, G.C.},
  \bibinfo{year}{2017}.
\newblock \bibinfo{title}{\uppercase{MAGE-A} antigens and cancer
  immunotherapy}.
\newblock \bibinfo{journal}{Front Med} \bibinfo{volume}{4},
  \bibinfo{pages}{18}.

\end{thebibliography}
\bibliographystyle{elsarticle-harv}
\setcounter{section}{0}
\setcounter{figure}{0}
\setcounter{table}{0}
\renewcommand{\thesection}{\Alph{section}}
\end{document}